\documentclass[aps,showpacs,amssymb,eqsecnum,twocolumn]{revtex4}
\usepackage{graphicx}
\usepackage{dcolumn}
\usepackage{bm}

\begin{document}

\title{Disorder and temperature dependence of the Anomalous Hall Effect in thin ferromagnetic films: Microscopic model}
\author{K. A. Muttalib}
\email{muttalib@phys.ufl.edu}
\affiliation{Department of Physics, University of Florida, P.O. Box 118440,
Gainesville, FL 32611-8440}
\author{P. W\"olfle}
\email{woelfle@tkm.uni-karlsruhe.de}
\affiliation{ITKM, Universit\"at Karlsruhe, D-76128 Karlsruhe, Germany,}
\affiliation{INT,  Forschungzentrum Karlsruhe, Postfach 3640, 76021 Karlsruhe, Germany}

\begin{abstract}
We consider the Anomalous Hall Effect (AHE) in thin disordered
ferromagnetic films. Using a microscopic model of electrons in a
random potential of identical impurities including spin-orbit
coupling, we develop a general formulation for strong, finite
range impurity scattering. Explicit calculations are done within a
short range but strong impurity scattering to obtain AH
conductivities for both the skew scattering and side jump
mechanisms. We also evaluate quantum corrections due to
interactions and weak localization effects. We show that for
arbitrary strength of the impurity scattering, the
electron-electron interaction correction to the AH conductivity
vanishes exactly due to general symmetry reasons. On the other
hand, we find that our explicit evaluation of the weak
localization corrections  within the strong, short range impurity
scattering model can explain the experimentally observed
logarithmic temperature dependences in disordered ferromagnetic Fe
films.

\end{abstract}


\maketitle

\section{Introduction}

It has been recognized since the 1950's \cite{KL} that a Hall
effect can exist in ferromagnetic metals even in the absence of an
external magnetic field, hence the name Anomalous Hall Effect
(AHE). There are several different mechanisms that might be
responsible for the AHE observed in thin ferromagnetic films,
namely the skew scattering \cite{smit} and side jump mechanisms
\cite{berger} as well as Berry phase contributions \cite{niu}. All
such mechanisms depend on spin-orbit interaction induced by the
impurities and on the spontaneous magnetization in a ferromagnet
which breaks the time reversal invariance and therefore gives rise
to the AHE. For a disordered ferromagnetic film, AH conductivity
due to the skew scattering and side jump mechanisms have been
theoretically considered using a variety of methods within weak,
short range impurity scattering
\cite{luttinger,lewiner,sinitsyn,dugaev,LW}. However, a systematic
calculation, starting from a microscopic Hamiltonian, of the
longitudinal as well as the AH conductivities for different
mechanisms for  strong impurity scattering has been lacking.
Recently, the effect of strong, short range impurity scattering on
the longitudinal and Hall conductivities were considered for skew
scattering as well as side jump mechanisms \cite{WM}, but quantum
corrections, namely electron-electron (e-e) interaction
corrections \cite {altshuler} or weak localization (WL) effects
\cite{lee}, were not included.

Earlier experiments \cite{BY} have shown logarithmic temperature
dependences of the longitudinal as well as Hall resistances
highlighting the importance of such quantum corrections. However,
the results were consistent with, and were interpreted as,
vanishing interaction contributions to the AH conductivity,
obtained theoretically within a {\it weak} impurity scattering
model \cite{LW} and the absence of any weak localization effects.
Recent experiments on the other hand clearly show non-vanishing
contribution to the total quantum correction to the AH
conductivity \cite{mitra}, which can arise in principle either
from an interaction correction due to {\it strong} impurity
scattering, or from a weak localization effect, or from a
combination of both. It has been commonly believed that weak
localization effects in ferromagnetic films would be cut off by
the presence of large internal magnetic field among others, which
suggests that the interaction corrections to the AH conductivity
need to be revisited for strong impurity scattering as a source of
difference between the two experiments.

In this paper we systematically develop a general formulation for
the AHE for {\it strong, finite range} impurity scatterings
starting from a microscopic model of electrons in a random
potential of impurities including spin-orbit coupling. This
generalizes an earlier work \cite{lewiner} which considered weak,
short range impurity scattering only and did not include quantum
corrections. We show on very general symmetry grounds that quantum
correction to the AH conductivity due to (e-e) interaction effects
vanish exactly, which shows that the previous weak scattering
results \cite{LW} remain valid for arbitrary strengths of the
impurity scattering. This forces us to consider the weak
localization effects \cite{dugaev} as the only remaining source of
the logarithmic temperature dependence in the above experiments
despite the presence of large internal magnetic fields and
spin-orbit scatterings in these ferromagnetic films. As we show
below, the temperature independent cutoff of the weak localization
effects in strongly disordered systems can be ineffective at
higher temperatures if a temperature dependent contribution
dominates the phase relaxation rate. It turns out that while the
contribution from the e-e interaction to the phase relaxation rate
is indeed too small for WL effects to be observed, a much larger
contribution is obtained from scattering off spin waves
\cite{tatara}, which should allow the observation of the WL
effects within a reasonable temperature range. We find that the
effects of strong impurity scatterings on the WL effects can be
evaluated to obtain a very simple result, namely that the ratio of
the WL corrections to the AH to the longitudinal conductivity can
be written simply in terms of the eigenvalues of the impurity
averaged particle-hole scattering amplitude for zero momentum
transfer. This result, taken together with contributions to the AH
conductivity from both the skew scattering and side jump
mechanisms calculated within the same microscopic model, can
explain both the earlier as well as the recent experiments on the
disorder and temperature dependences of the AH conductivities of
ultrathin Fe films \cite{mitra} mentioned above. This last  result
has been reported without details in combination with the recent
experiment in a short letter \cite{mitra}.

The paper is organized in the following way: A microscopic model
Hamiltonian is introduced in Section II, and a general formulation
in two dimensions for strong, finite range impurity scatterings is
developed in Section III. Section IV reviews the results on the
conductivity tensor in the absence of interactions. In Sections V
and VI we consider the e-e interaction corrections and the weak
localization corrections, respectively, to both longitudinal and
AH conductivities within the general strong, finite range impurity
scattering formulation. We then consider the special case of a
short range, but still strong, impurity scattering model in
Section VII. In Section VIII we collect all the results and
compare them with recent experiments. Section IX summarizes the
paper. For the sake of completeness, we include models of small
and large angle scatterings in the Appendix.

\section{Hamiltonian}

The single particle Hamiltonian of a conduction electron in a
ferromagnetic disordered metal, including spin-orbit interaction
induced by the disorder potential $V_{dis}(\mathbf{r)}$, is given
in its simplest form by  (throughout the paper we use units with
$\hbar = k_B= 1$)
\begin{eqnarray}
H_{1}&=&[-\frac{\nabla ^{2}}{2m}+V_{dis}(\mathbf{r)]\delta
}_{\sigma \sigma \prime }-M\tau _{\sigma \sigma \prime }^{z}\cr
&-& i\frac{\lambda _{c}^{2}}{(4\pi )^{2}}[\mathbf{\tau }_{\sigma
\sigma \prime }\cdot (\nabla V_{dis}\times \nabla )],
\end{eqnarray}
where $\lambda _{c}=\frac{2\pi }{mc}$ is the Compton wavelength of
the electron, and $M$ is the Zeeman energy splitting caused by the
ferromagnetic polarization. Here $H_1$ is a $2\times 2$ matrix in
spin space with $\sigma, \sigma' = \uparrow,\downarrow$ being spin
indices and $\tau$ is the vector of Pauli matrices. The above
model is only a crude approximation of the bandstructure of Fe,
which has been determined by several authors (see e.g. ref
\cite{singh}). We model the energy band crossing the Fermi surface
by a single isotropic band. As will be discussed below, the
quantum corrections to the conductivity exhibit certain
qualitative features, which do not depend sensitively on the
details of the band structure. The disordered potential in (2.1)
will be modelled as randomly placed identical impurities,
$V_{dis}(\mathbf{r})=\sum_{j}V(\mathbf{r-R}_{j})$. We will later
average over the impurity positions $\mathbf{R}_{j}$.

The matrix elements of  $H_{1}$ in the plane wave (or Bloch state)
representation are given by
\begin{eqnarray}
&\;&\langle \mathbf{k\prime \sigma \prime |}H_{1}|\mathbf{k}\sigma
\rangle = \int d^{2}r e^{ -i\mathbf{k\prime \cdot r}}H_{1}e^{ -i\mathbf{k\cdot r}} \cr &=&
(\frac{k^{2}}{2m}-M\sigma )\mathbf{\delta }_{\mathbf{kk}
\prime }\mathbf{\delta }_{\sigma \sigma \prime }+ \sum_j
V(\mathbf{k-k\prime )}e^{ i(\mathbf{k-k\prime )\cdot } \mathbf{R}_{j}} \cr
&+&V_{so}(\mathbf{k\prime \sigma \prime ;k\sigma )}
\end{eqnarray}
where $V(\mathbf{k-k\prime )}$ is the Fourier transform of the single
impurity potential, and the spin-orbit interaction part is given by
\begin{eqnarray}
V_{so}(\mathbf{k\prime \sigma \prime ;k\sigma )}
&=&-i\frac{\lambda _{c}^{2}}{(4\pi )^{2}}\sum_jV(\mathbf{k-k\prime
)}e^{ [i(\mathbf{k-k\prime )\cdot }\mathbf{R}_{j}]}\cr &\times&
\mathbf{\tau }_{\sigma \sigma \prime }\cdot (\mathbf{k\times
k\prime )}
\end{eqnarray}
Here we have used
\begin{eqnarray}
&\;& -i\int d^{2}r \exp (-i\mathbf{k\prime \cdot r})(\nabla
V_{dis}\times \nabla )\exp (-i\mathbf{k\cdot r}) \cr
&=&\mathbf{-}i\int d^{2}r\int \frac{d^{2}q}{
(2\pi )^{2}}e^{ i(\mathbf{k-k\prime -q)\cdot } \mathbf{r}}(-i\mathbf{
q)V(q)\times (}i\mathbf{k)} \cr
&\;&\cr
&=& -iV(\mathbf{k-k\prime )(k\times
k\prime )}
\end{eqnarray}
The many-body Hamiltonian is given in terms of electron creation and
annihilation operators $c_{\mathbf{k\sigma }}^{+},c_{\mathbf{k\sigma }}$ as
\begin{eqnarray}
H&=&\sum_{\mathbf{k\sigma }}{ }(\varepsilon _{\mathbf{k}}-M\sigma
)c_{\mathbf{k\sigma }}^{+}c_{\mathbf{k\sigma }}\cr
&+&\sum_{\mathbf{k\sigma ,k\prime \sigma \prime }}
\sum_{j}V(\mathbf{k-k\prime )}e^{ i(\mathbf{k-k\prime )\cdot
}\mathbf{R}_{j}}\cr &\times& \{\mathbf{\delta }_{\sigma \sigma
\prime }-i \bar{g} _{so}\mathbf{\tau }_{\sigma \sigma \prime
}\cdot ( \widehat{k}\mathbf{\times} \widehat{k}^{\prime}
)\}c_{\mathbf{k\prime \sigma \prime }}^{+}c_{\mathbf{k\sigma }}
\end{eqnarray}
where we have defined a dimensionless spin-orbit coupling constant
$\bar{g} _{so}\equiv \frac{\lambda _{c}^{2}k_{F}^{2}}{(4\pi
)^{2}}$, $\widehat{k}\equiv \mathbf{k/|k|}$. Note: An estimate of
the spin-orbit coupling constant $\bar{g} _{so}$, using a typical
Fermi wave number $k_{F\text{ }}$, shows that it is rather small,
of order $10^{-4}$. However, in transition metal compounds the
coupling is substantially enhanced by interband mixing effects
\cite{berger}, so that the renormalized coupling constant  $g
_{so}$ is of order unity: $g _{so}\thicksim c_{so}E_{so}/\Delta
E_{d}$, where $E_{so}\thicksim 0.1eV$ \ is a measure for the
atomic spin-orbit energy, \ $\Delta E_{d}\thicksim 0.5eV$ \ is a
typical energy splitting of d-bands, and the constant $\ c_{so}$
$\thicksim 5$. In the following we will replace $\bar{g} _{so}$ by
the phenomenological spin-dependent parameter $g_{\sigma}$.

\section{Impurity scattering: General Formulation}

In this section, we will develop a general formulation for {\it
strong, finite range} impurity scattering in two dimensions using
standard field theory techniques at finite temperature \cite{AGD}.
For simplicity, we will need to make approximations for short
range impurity scattering later. However, keeping the formulation
general as long as possible will allow us e.g. to check if the
anisotropic scattering can have a large impact on our final
results.

The repeated scattering of an electron off a single impurity may
be described symbolically in terms of the scattering amplitude
$f_{k\sigma ,k\prime \sigma \prime }$ as
\begin{eqnarray}
f=V+VGV+VGVGV+....
\end{eqnarray}
where $G$ is the single particle Green's function
\begin{equation}
G_{k\sigma }(i\omega _{n})=[i\omega _{n}-\varepsilon _{k\sigma }-\Sigma
_{k\sigma }(\omega _{n})]^{-1},
\end{equation}
with the single particle self energy  $\Sigma _{k\sigma }(i\omega
_{n})$. Here $\omega _{n}=\pi T(2n+1)$ is the fermion Matsubara
frequency with $T$ being the temperature and $N_{\sigma}$ is the
density of states at the Fermi level of spin species $\sigma$. (We
use units of temperature such that Boltzmann's constant is equal
to unity). V is the bare interaction with one impurity at
$\mathbf{R=0}$ and includes the spin-orbit scattering
\begin{equation}
V_{k,k\prime; \sigma }=V(k-k^{\prime })[1-ig _{\sigma}\tau
_{\sigma \sigma }^{z}(\widehat{k}\times \widehat{k}^{\prime} )],
\end{equation}
where we have used the fact that $V$ is diagonal in spin space. In
the case of finite range, or even long-range correlated scattering
potentials, we may still use the model of individual impurities or
scattering centers, but now of finite spatial extension. This is
reasonable as long as the scattering centers do not overlap too
much. If they overlap, a more statistical description in terms of
correlators of the impurity potential should be used. Within our
model, the nonlocal character of scattering is described in terms
of the momentum dependence of the Fourier Transform of the
potential of a single impurity (assuming only one type of
impurity) $V(\mathbf{k-k\prime )}$, which for an isotropic system
depends only on the angle $\theta $ between $\mathbf{k}$ and
$\mathbf{k^{\prime} }$, \ $V=V(\theta )=V(-\theta ).$ In 2d we may
expand $V$ in terms of eigenfunctions
$\chi_{m}(\widehat{k})=e^{im\phi}$, where $\phi$ is the polar
angle of vector $\mathbf{k}$,
$\widehat{k}=\mathbf{k}/|\mathbf{k}|$.  Adding the skew scattering
potential we may write
\begin{equation}
V_{k,k\prime \sigma }=\sum_{m}V_{m\sigma } \chi
_{m}(\widehat{k})\chi _{m}^{\ast } (\widehat{k}^{\prime} )
\end{equation}
where $V_{m\sigma}$ is a sum of the normal and skew scattering parts
\begin{equation}
V_{m\sigma }=V_{m}^{ns}+V_{m\sigma }^{ss}.
\end{equation}
Time reversal invariance and rotation symmetry in  the case of
potential scattering implies
\begin{equation}
V_{-m}^{ns}=(V_{m}^{ns})^{\ast }=V_{m}^{ns} .
\end{equation}
Equation (3.3) then yields
\begin{equation}
V_{m\sigma }^{ss}=\frac{1}{2}g _{\sigma}\tau^z_{\sigma\sigma}
(V^{ns} _{m-1}-V^{ns} _{m+1})
\end{equation}

\subsection{Scattering amplitude}

For $V$ diagonal in spin space, the scattering amplitude \
$f_{k\sigma ,k\prime \sigma \prime }=\delta _{\sigma ,\sigma
^{\prime }}f_{k,k\prime \sigma }$ obeys the integral equation
\begin{eqnarray}
f_{k,k\prime \sigma }^{s} &=& V_{k,k\prime \sigma }
+\sum_{k_{1}}{}  G_{k_{1}\sigma }(i\omega _{n})V_{k,k_{1}\sigma
}f_{k_{1},k\prime \sigma}^{s} \cr &=& V_{k,k\prime \sigma }-is\pi
N_{\sigma }\langle V_{k,k_{1}\sigma }f_{k_{1},k\prime \sigma
}^{s}\rangle _{k_{1}} ,
\end{eqnarray}
where $s \equiv sign(\omega _{n})$ and $\langle \cdots
\rangle_{k_1}$  denotes averaging over the direction of wavevector
$\mathbf{k_1}$. Defining  dimensionless potential
$\bar{V}_{m\sigma} \equiv \pi N_{\sigma }V_{m\sigma }$ and the
dimensionless scattering amplitude $\bar{f}_{k\sigma ,k\prime
\sigma \prime }\equiv \pi N_{\sigma }f_{k\sigma ,k\prime \sigma
\prime }$ and expanding $\bar{f}_{k\sigma ,k\prime \sigma
}=\sum_{m}\bar{f}_{m\sigma }\chi _{m}(\widehat{k})\chi _{m}^{\ast
}(\widehat{k}^{\prime} )$, we find
\begin{equation}
\bar{f}^s_{m\sigma}=\frac{\bar{V}_{m\sigma}}{1+is
\bar{V}_{m\sigma}} .
\end{equation}

For notational simplicity, we will always use a bar on a symbol to
represent the corresponding dimensionless quantity.

\subsection{Single particle relaxation rate}

The single particle relaxation rate $\tau_{\sigma}$ is given by
the imaginary part of the self energy
\begin{eqnarray}
\frac{1}{2\tau _{\sigma }}&\equiv &-sIm\Sigma _{k\sigma }(i\omega
_{n})\cr &=&-sn_{imp}Im (f_{k\sigma ,k\sigma }^{s})=
\frac{n_{imp}}{\pi N_{\sigma }}\gamma _{\sigma }
\end{eqnarray}
where $\gamma _{\sigma }$ is a dimensionless parameter
characterizing the scattering strength, $\gamma _{\sigma }\equiv
-s\sum_{m}Im (\bar{f}_{m\sigma })=
\sum_{m}\frac{\bar{V}^2_{m\sigma }}{1+\bar{V}^2_{m\sigma }}$, and
$N_{\sigma}$ is the density of states at the Fermi energy of spin
species $\sigma$. Note that $\bar{V}_{m\sigma}$ are all real.

\subsection{Particle-hole propagator}

The particle-hole propagator $\Gamma _{kk\prime }(q;i\epsilon
_{n},i\epsilon _{n}-i\Omega _{m})$ \ is an important ingredient of
vertex corrections of any kind. Here $k+q/2,k-q/2$ \ are the
initial , $k\prime +q/2,k\prime -q/2$ the final momenta and \
$\epsilon _{n},\epsilon _{n}-\Omega _{m}$ \ are the Matsubara
frequencies of the particle and the hole line, respectively. In
terms of the particle-hole scattering amplitude
$t_{k,k\prime}(q;i\epsilon _{n},i\Omega _{m})$, $ \Gamma $
satisfies the following Bethe-Salpeter equation (we have defined
dimensionless quantities $\bar{\Gamma},\bar{t}$  by multiplying
both with a factor $(2\pi N_{\sigma }\tau _{\sigma })$)
\begin{eqnarray}
&\;&\bar{\Gamma}_{kk\prime }(q;i\epsilon _{n},i\Omega _{m})=
\bar{t}_{kk\prime }(q;i\epsilon _{n},i\Omega _{m})\cr &\;&\cr
&+&(2\pi N_{\sigma}\tau _{\sigma })^{-1}
\sum_{k_{1}}\bar{t}_{kk_{1}}(q;i \epsilon _{n}, i\Omega _{m})
G_{k_{1}+q/2,\sigma }(i\epsilon_{n}) \cr &\times&
G_{k_{1}-q/2,\sigma }(i\epsilon _{n}-i\Omega _{m})\bar{\Gamma}
_{k_{1}k\prime }(q;i\epsilon _{n},i\Omega _{m})
\end{eqnarray}
The (dimensionless) impurity averaged particle-hole scattering
amplitude $\bar{t}$ (we consider only the case of equal spin of
particle and hole) is given in terms of the (dimensionless)
scattering amplitudes $\bar{f}$ by the equation
\begin{eqnarray}
&\;&\bar{t}_{kk\prime }^{ss\prime }(q;i\epsilon _{n},i\Omega _{m})
=\frac{2\tau _{\sigma }n_{imp}}{\pi N_{\sigma
}}\bar{f}_{k+q/2,\sigma;k\prime +q/2\sigma }^{s}(i\epsilon
_{n})\cr &\times&\bar{f}_{k\prime -q/2,\sigma ;k-q/2,\sigma
}^{s\prime }(i\epsilon _{n}-i\Omega _{m}).
\end{eqnarray}
We will later need the limit of small $q,$ $q\ll k_{F},$  of this
expression,
\begin{equation}
\bar{t}_{kk\prime }^{ss\prime }(q;i\epsilon _{n},i\Omega _{m})=
\bar{t}_{kk\prime }^{ss\prime }(q=0)+\Delta \ \bar{t}_{kk\prime
}^{ss\prime }(q).
\end{equation}
It is useful to represent the operator $\bar{t}_{kk\prime }(q=0)$
\ in terms of its eigenvalues $\lambda _{m}$. Assuming isotropic
band structure, the eigenfunctions $\chi _{m}(\widehat{k})=\exp
(im\varphi )$  are those of the angular momentum operator
component $L_{z}$.   The eigenvalue equation is
\begin{equation}
\langle \bar{t}_{kk\prime }(q=0) \chi _{m}(\widehat{k}^{\prime} )
\rangle _{k\prime } =\lambda _{m}\chi _{m}(\widehat{k}).
\end{equation}
The operator $\ \bar{t}_{kk\prime }^{+-}(q=0)$   may be
represented as
\begin{eqnarray}
\bar{t}_{kk\prime }^{+-}(q=0)&=&\sum_{m}\lambda _{m}\chi
_{m}(\widehat{k})\chi _{m}^{\ast }(\widehat{k} ^{\prime})\cr
\bar{t}_{kk\prime }^{-+}(q=0)&=&[\ \bar{t} _{k\prime
k}^{+-}(q=0)]^{\ast }
\end{eqnarray}
In general, using the definitions
\begin{eqnarray}
t_{k,k\prime \sigma }^{ss^{\prime }}&=& \frac{n_{imp}}{(\pi
N_{\sigma })^{2}} \bar{f}_{k,k^{\prime }\sigma
}^{s}\bar{f}_{k^{\prime },k\sigma }^{s^{\prime }}=(2\pi N_{\sigma
}\tau _{\sigma })^{-1}\bar{t} _{k,k^{\prime }\sigma }^{ss^{\prime
}}\cr \bar{t} _{k,k^{\prime }\sigma }^{ss^{\prime
}}&=&\sum_{m}\bar{t} _{m\sigma }^{ss^{\prime }}\chi
_{m}(\widehat{k}) \chi _{m}^{\ast }(\widehat{k}^{\prime} )
\end{eqnarray}
we have
\begin{equation}
\bar{t}_{m\sigma }^{ss^{\prime }} =\gamma _{\sigma
}^{-1}\sum_{m\prime } \bar{f}_{m\prime \sigma
}^{s}\bar{f}_{m\prime -m,\sigma }^{s^{\prime }}.
\end{equation}
We will consider $\Delta \ \bar{t}_{kk\prime }(q)$ for the special
case of strong short range impurity scatterings in section VII-A.

The energy integral over the product of Green's functions in the
integral equation for $\Gamma _{kk\prime }$ may be done first,
after expanding the G's in \ $\Omega _{m}$ and  $q$,
\begin{eqnarray}
\int &d\varepsilon _{1}& G_{k_{1}+q/2,\sigma }(i\epsilon
_{n})G_{k_{1}-q/2,\sigma }(i\epsilon _{n}-i\Omega _{m})\cr
&=& 2\pi \tau \lbrack 1+i\tau (i\Omega _{m}-\mathbf{q\cdot v}_{k_{1}})\cr
&-&\tau
^{2}(\mathbf{q\cdot v}_{k_{1}})^{2}]
\end{eqnarray}
with $\epsilon _{n}>0$ and $\epsilon_{n}-\Omega _{m}<0$, where
$\mathbf{q\cdot v}_{k}=qv_{F}(\widehat{q}\cdot \widehat{k})$.
Expanding $\bar{\Gamma}_{kk\prime }$ and $\bar{t}_{kk\prime }$  in
terms of eigenfunctions $\chi _{m}(\widehat{k})$,
$\bar{\Gamma}_{kk\prime }=$\ $\sum_{m}\bar{\Gamma}_{mm\prime }\chi
_{m}(\widehat{k})\chi _{m^{\prime }}^{\ast }(\widehat{k}^{\prime}
)$ and using $\tilde{t}^{+-}_{m\sigma} \equiv \lambda_m$  one
obtains ($s^{\prime }=-s$)
\begin{eqnarray}
\bar{\Gamma}_{mm\prime }^{ss\prime } &=&\lambda _{m}\delta
_{mm^{\prime }}+\lambda _{m}\{[1-\tau (|\Omega
_{n}|+D_{0}q^{2})]\bar{\Gamma} _{mm\prime }^{ss^{\prime }}\cr
&-&\frac{i}{2}v_{F}q\tau  s[\bar{\Gamma}_{m-1,m\prime
}^{ss^{\prime }}\chi _{1}^{\ast }(\widehat{q}) +
\bar{\Gamma}_{m+1,m\prime }^{ss^{\prime }}\chi
_{1}(\widehat{q})]\cr &-&\frac{1}{4}(v_{F}q\tau
)^{2}[\bar{\Gamma}_{m-2,m\prime }^{ss^{\prime }}\chi _{2}^{\ast
}(\widehat{q})\cr &+&\bar{\Gamma}_{m+2,m\prime }^{ss^{\prime
}}\chi _{2}(\widehat{q})]\}
\end{eqnarray}
For \ $m=m^{\prime }\neq 0$ \ the solution is
\begin{equation}
\bar{\Gamma}_{mm}= \frac{\lambda _{m}}{1-\lambda _{m}}+O(q) \equiv
\widetilde{\lambda}_m + O(q),
\end{equation}
where we have defined $\widetilde{\lambda}_m \equiv
\lambda_m/(1-\lambda_m)$. The $\lambda_m$ (and therefore
$\widetilde{\lambda}_m$) are complex valued and depend on the spin
projection $\sigma$. Using conventional notation, we will denote
the real and imaginary parts of $\lambda_m$ by
$\lambda^{\prime}_m$ and $\lambda^{\prime\prime}_m$, respectively,
and similarly the real and imaginary parts of
$\widetilde{\lambda}_m$ by $\widetilde{\lambda}^{\prime}_m$ and
$\widetilde{\lambda}^{\prime\prime}_m$, respectively.

The case $m=0$ needs special consideration, because particle
number conservation causes $\bar{\Gamma}_{00}$ to have a pole in
the limit $\Omega _{n},q\rightarrow 0$, here expressed by $\lambda
_{0}=1$. Solving the above equation for $\bar{\Gamma}_{00}$ \ in
lowest order in q, one finds
\begin{equation}
\bar{\Gamma}_{00}=\frac{1/\tau }{|\Omega _{m}|+Dq^{2}} ,
\end{equation}
where the renormalized diffusion constant is defined as
\begin{eqnarray}
D&=&D_{0}(1+\ \widetilde{\lambda }_{1}^{\prime }) ,   \;\;\;
D_0=\frac{1}{2}v^2_F\tau \cr \widetilde{ \lambda }_{1}^{\prime
}&\equiv& \rm{Re}\widetilde{\lambda }_{1} =
\frac{1}{2}(\widetilde{\lambda }_{1}+\widetilde{ \lambda }_{-1});
\end{eqnarray}
This is found by solving the following equations for small $v_{F}q\tau (
s^{\prime }=-s)$
\begin{eqnarray}
\bar{\Gamma}_{00}^{ss\prime }&=&1+[1-\tau (|\Omega
_{m}|+D_{0}q^{2})] \bar{\Gamma}_{00}^{ss^{\prime }}\cr
&-&\frac{i}{2}v_{F}q\tau s[\widetilde{ \Gamma }_{-1,0}^{ss^{\prime
}}\chi _{1}^{\ast }(\widehat{q})+ \bar{\Gamma}_{1,0}^{ss^{\prime
}}\chi _{1}(\widehat{q})]\cr \bar{\Gamma}_{-1,0}^{ss\prime }&=&
\lambda _{-1}\{\bar{\Gamma}_{-1,0}^{ss^{\prime }}
-\frac{i}{2}v_{F}q\tau s\bar{\Gamma}_{0,0}^{ss^{\prime }}\chi
_{1}(\widehat{q})\}\cr \bar{\Gamma}_{1,0}^{ss\prime } &=&\lambda
_{1}\{\bar{\Gamma} _{1,0}^{ss^{\prime }}-\frac{i}{2}v_{F}q\tau
s\bar{\Gamma} _{0,0}^{ss^{\prime }}\chi _{-1}(\widehat{q})\}
\end{eqnarray}
Substituting $\bar{\Gamma}_{\pm 1,0}^{ss\prime }$ into the
equation for $\bar{\Gamma}_{00}^{ss\prime }$ one finds:
\begin{equation}
\bar{\Gamma}_{00}^{ss\prime }\{|\Omega
_{m}|+D_{0}q^{2}[1+\frac{1}{2} (\widetilde{\lambda
}_{1}+\widetilde{\lambda }_{-1})]\}=\frac{1}{\tau }
\end{equation}
The leading singular dependence on $\widehat{k}\prime $ is obtained
from:
\begin{eqnarray}
\bar{\Gamma}_{0,\pm 1}^{ss\prime }&=& [1-\tau (|\Omega
_{m}|+D_{0}q^{2})]\bar{\Gamma}_{0,\pm 1}^{ss^{\prime }}\cr
&-&\frac{i}{2} v_{F}q\tau s[\bar{\Gamma}_{-1,\pm 1}^{ss^{\prime
}}\chi _{1}^{\ast }( \widehat{q})+\bar{\Gamma}_{1,\pm
1}^{ss^{\prime }}\chi _{1}(\widehat{q})]\cr
\bar{\Gamma}_{-1,1}^{ss\prime } &=& -\frac{i}{2}\widetilde{\lambda
} _{-1}v_{F}q\tau s\bar{\Gamma}_{0,1}^{ss^{\prime }}\chi
_{1}(\widehat{q})\cr \bar{\Gamma}_{1,1}^{ss\prime
}&=&\widetilde{\lambda }_{1}-\frac{i}{2} \widetilde{\lambda
}_{1}v_{F}q\tau s\bar{\Gamma}_{0,1}^{ss^{\prime }}\chi _{1}^{\ast
}(\widehat{q})
\end{eqnarray}

The complete particle-hole propagator in the regime $v_{F}q\tau <1$ is given by
\begin{equation}
\bar{\Gamma}_{kk\prime }=\frac{1}{\tau }\frac{\gamma
_{k}\widetilde{\gamma }_{k\prime }}{|\Omega _{m}|+Dq^{2}}+\
\sum_{m\neq 0}\widetilde{ \lambda }_{m}\chi _{m}(\widehat{k})\chi
_{m}^{\ast }(\widehat{k}^{\prime} )
\end{equation}
with
\begin{eqnarray}
\gamma _{k}&=&1-\frac{i}{2}v_{F}q\tau s\sum_{m=\pm 1}
\widetilde{\lambda }_{m}\chi _{m}(\widehat{k})\chi {m}^{\ast }(\widehat{q})\cr
&=&1-\frac{i}{2}v_{F}\tau s\sum_{m=\pm 1}\widetilde{\lambda }
_{m}\chi _{m}(\widehat{k})q_{-m}
\end{eqnarray}
and
\begin{equation}
\widetilde{\gamma }_{k}=1-\frac{i}{2}v_{F}\tau s\sum_{m=\pm 1}
\widetilde{\lambda }_{m}\chi _{m}^{\ast }(\widehat{k})q_m.
\end{equation}
The vertex corrections of the density $T_{k}$  and current
vertices  $j_{k\alpha }$  and $\widetilde{j}_{k\alpha }$ (for the
incoming and outgoing current) are obtained by
\begin{equation}
T_{k}(q)\equiv 1+\langle \bar{\Gamma}_{kk\prime }\rangle _{k\prime
}=1+\frac{1/\tau }{|\Omega _{m}|+Dq^{2}}\gamma _{k}
\end{equation}
and
\begin{eqnarray}
j_{k\alpha }(q) &=& v_{k\alpha } +\langle v_{k\prime\alpha
}\bar{\Gamma}_{k\prime k} \rangle _{k\prime }  \cr &\;&\cr &=&
v_{k\alpha }+\sum_{m=\pm 1} \widetilde{\lambda }_{m}\chi
_{m}^{\ast }(\widehat{k} )\langle v_{k\prime \alpha }\chi
_{m}(\widehat{k}^{\prime})\rangle _{k\prime }\cr &+&\langle
 v_{k\prime \alpha }\gamma _{k\prime
}\rangle _{k\prime }\frac{1/\tau }{|\Omega
_{m}|+Dq^{2}}\widetilde{\gamma }_{k}\cr \widetilde{j}_{k\alpha
}(q) &=&v_{k\alpha } +\langle v_{k\prime \alpha
}\bar{\Gamma}_{kk\prime }\rangle _{k\prime } \cr &\;&\cr &=&
v_{k\alpha } +\sum_{m=\pm 1}\widetilde{\lambda }_{m}\chi_{m}
(\widehat{k})\langle v_{k\prime \alpha }\chi _{m}^{\ast
}(\widehat{k}^{\prime} )\rangle _{k\prime }\cr &+&\langle
v_{k\prime \alpha } \widetilde{\gamma }_{k}\rangle _{k\prime
}\frac{1/\tau }{|\Omega _{m}|+Dq^{2}}\gamma _{k}.
\end{eqnarray}
Note that $\widetilde{j}_{k\alpha }\neq ($\ $j_{k\alpha })^{\ast
}$,  as the eigenvalues $\widetilde{\lambda }_{m}$ are in general
complex valued. Using
\begin{eqnarray}
\chi _{-1}(\widehat{k})\chi _{1}^{\ast }(\widehat{k}^{\prime} )
+\chi _{1}(\widehat{k})\chi _{-1}^{\ast }(\widehat{k}^{\prime}
)&=& 2(\widehat{k}\mathbf{\cdot }\widehat{k} ^{\prime} )\cr \chi
_{-1}(\widehat{k})\chi _{1}^{\ast }(\widehat{k}^{ \prime} )-\chi
_{1}(\widehat{k})\chi _{-1}^{\ast}(\widehat{k}^{\prime} )&=&
2i(\widehat{k} \mathbf{\times}\widehat{k}^{\prime})
\end{eqnarray}
and
\begin{equation}
\left\langle \widehat{k}_{\alpha }^{\prime }(\widehat{k}\mathbf{
\cdot }\widehat{k}^{\prime} )\right\rangle =\frac{1}{2}\widehat{k}
_{\alpha }; \;\;\;
\left\langle \widehat{k}_{\alpha }^{\prime }(\widehat{k}\mathbf{
\times }\widehat{k}^{\prime} )\right\rangle =-\frac{1}{2}(\widehat{
e}_{\alpha }\times \widehat{k})
\end{equation}
and defining $j_{k\alpha }\equiv j_{k\alpha }(q=0)$,
$\widetilde{j}_{k\alpha }\equiv \widetilde{j}_{k\alpha }(q=0)$, we
have
\begin{eqnarray}
j_{k\alpha }&=&v_{F}[(1+\ \widetilde{\lambda }_{1}^{\prime
})\widehat{k }_{\alpha }+\ \widetilde{\lambda }_{1}^{\prime \prime
}(\widehat{e} _{\alpha }\times \widehat{k})_{z}]\cr
\widetilde{j}_{k\alpha }&=&v_{F}[(1+\ \widetilde{\lambda
}_{1}^{\prime }) \widehat{k}_{\alpha }-\ \widetilde{\lambda
}_{1}^{\prime \prime } (\widehat{e}_{\alpha }\times
\widehat{k})_{z}].
\end{eqnarray}
More explicitly, for $\alpha=x,y$, the incoming and outgoing
current vertices $j$ and $\widetilde{j}$ have the forms
\begin{eqnarray}
j_{kx}&=& v_{F}[(1+\ \widetilde{\lambda }_{1}^{\prime
})\widehat{k} _{x}+\ \widetilde{\lambda }_{1}^{\prime \prime
}\widehat{k}_{y}]\cr &=&\frac{1}{2} v_{F}[(1+\ \widetilde{\lambda
}_{1}^{\ast })\widehat{k}_{+}+(1+\ \widetilde{ \lambda
}_{1})\widehat{k}_{-}]\cr j_{ky}&=& v_{F}[(1+\ \widetilde{\lambda
}_{1}^{\prime }) \widehat{k}_{y}-\ \widetilde{\lambda
}_{1}^{\prime \prime }\widehat{k} _{x}]\cr
&=&-i\frac{1}{2}v_{F}[(1+\ \widetilde{\lambda }_{1}^{\ast
})\widehat{k} _{+}-(1+\ \widetilde{\lambda
}_{1})\widehat{k}_{-}]\cr \widetilde{j}_{kx}&=& v_{F}[(1+
\widetilde{\lambda  }_{1}^{\prime })\widehat{k}_{x}-\
\widetilde{\lambda }_{1}^{\prime \prime }\widehat{k}_{y}]\cr &=&
\frac{1}{2}v_{F}[(1+\ \widetilde{\lambda }_{1})\widehat{k}_{+}+(1+
\widetilde{\lambda }_{1}^{\ast })\widehat{k}_{-}]\cr
\widetilde{j}_{ky}&=& v_{F}[(1+\ \widetilde{\lambda } _{1}^{\prime
})\widehat{k}_{y}+\ \widetilde{\lambda }_{1}^{\prime \prime }
\widehat{k}_{x}]\cr &=&-i\frac{1}{2}v_{F}[(1+\ \widetilde{\lambda
}_{1})\widehat{k} _{+}-(1+\ \widetilde{\lambda }_{1}^{\ast
})\widehat{k}_{-}]
\end{eqnarray}
where we have defined $k_{\pm}=k_x\pm i k_y$.

\subsection{Particle-particle propagator}

The integral equation for the particle-particle propagator or
Cooperon reads (again multiplying the Cooperon $C$ and the
particle-particle scattering amplitude $t^{p}$ by the factor $2\pi
N_{\sigma }\tau _{\sigma }$ to define dimensionless Cooperon
$\bar{C}$ and dimensionless particle-particle scattering amplitude
$\bar{t^p}$)
\begin{eqnarray}
&\;&\bar{C}_{kk\prime }(Q ;i\epsilon _{n},i\Omega _{m})=
\bar{t}_{kk\prime }^{p}(Q;i\epsilon _{n},i\Omega _{m})\cr &\;&\cr
&+&(2\pi N_{\sigma }\tau _{\sigma })^{-1} \sum_{k_{1}}
\bar{t}_{kk_{1}}^{p}(Q;i \epsilon _{n},i\Omega _{m})
G_{k_{1},\sigma }(i\epsilon _{n})\cr &\times& G_{Q-k_{1},\sigma
}(i\epsilon _{n}-i\Omega _{m})\bar{C} _{k_{1}k\prime }(q;i\epsilon
_{n},i\Omega _{m})
\end{eqnarray}
\begin{eqnarray}
t_{k,k\prime \sigma }^{p,ss^{\prime }}&=&\frac{n_{imp}}{(\pi
N_{\sigma })^{2}} \bar{f}_{k,k^{\prime }\sigma
}^{s}\bar{f}_{-k,-k^{\prime },\sigma }^{s^{\prime }}\cr &=&(2\pi
N_{\sigma }\tau _{\sigma })^{-1}\gamma _{\sigma
}^{-1}\bar{f}_{k,k^{\prime }\sigma }^{s}\bar{f} _{-k,-k^{\prime
},\sigma }^{s^{\prime }}
\end{eqnarray}
\begin{equation}
\bar{t}_{k,k^{\prime }\sigma }^{p,ss^{\prime }}=2\pi N_{\sigma
}\tau _{\sigma }t_{k,k\prime \sigma }^{p,ss^{\prime }}=\sum_{m}
\bar{t}_{m\sigma }^{p,ss^{\prime }}\chi _{m}(\widehat{k})\chi
_{m}^{\ast }(\widehat{k}^{\prime} )
\end{equation}
\begin{equation}
\bar{t}_{m\sigma }^{p,ss^{\prime }}=\gamma _{\sigma }^{-1}
\sum_{m\prime }\bar{f}_{m\prime \sigma }^{s}\bar{f} _{m-m\prime
,\sigma }^{s^{\prime }}
\end{equation}
If rotation invariance or time reversal invariance is broken, \
$\bar{t }_{0\sigma }^{p,ss^{\prime }}=\gamma _{\sigma }^{-1}\gamma
_{\sigma }^{p}\neq 1$, where   $\gamma _{\sigma
}^{p}=\sum_{m\prime }\bar{f}_{m\prime \sigma
}^{s}\bar{f}_{-m\prime ,\sigma }^{s^{\prime }}$.

The energy integral over the product of Green's functions in the
integral equation for $C_{kk\prime }$ may be done first,  after
expanding the $G$'s in  $\Omega _{m}$ and  $Q$ ,
\begin{eqnarray}
\int &d\varepsilon _{1}& G_{k_{1},\sigma }(i\epsilon
_{n})G_{Q-k_{1},\sigma }(i\epsilon _{n}-i\Omega _{m}) \cr &=&
2\pi \tau \lbrack 1+i\tau (i\Omega _{m}
-\mathbf{Q\cdot v}_{k_{1}})\cr
&-&\tau
^{2}(\mathbf{Q\cdot v}_{k_{1}})^{2}],
\end{eqnarray}
with $\epsilon _{n}>0$, $\epsilon _{n}-\Omega _{m}<0$,   where \
$\mathbf{Q\cdot v}_{k}=Qv_{F}(\widehat{Q}\cdot \widehat{k})$.
Expanding $\bar{C}_{kk\prime }$  and  $ \bar{t}_{kk\prime }^{p}$
in terms of eigenfunctions $\chi _{m}(\widehat{k})$,
$\bar{C}_{kk\prime }=\sum_{m}\bar{C} _{mm\prime }\chi
_{m}(\widehat{k})\chi _{m^{\prime }}^{\ast }(\widehat{k}^{\prime}
)$ and denoting $\tilde{t}^{p,+-}_{m\sigma}=\lambda^p_m$ one
obtains ($s^{\prime }=-s$)
\begin{eqnarray}
&\;&\bar{C}_{mm\prime }=\lambda _{m}^{p}\{\delta _{mm^{\prime
}}+[1-\tau (|\Omega _{n}|+D_{0}Q^{2})]\bar{C}_{mm\prime }\cr
&-&\frac{i}{2}v_{F}Q\tau \lbrack \bar{C}_{m-1,m\prime }\chi
_{1}^{\ast }(\widehat{Q})
 + \bar{C}_{m+1,m\prime }\chi _{1}(
\widehat{Q})]\cr &-&\frac{1}{4}(v_{F}Q\tau )^{2}[\bar{C}
_{m-2,m\prime }\chi _{2}^{\ast }(\widehat{Q})\cr&+&\bar{C}
_{m+2,m\prime }\chi _{2}(\widehat{Q})]\}
\end{eqnarray}
The $m=m^{\prime }=0$ component of \ $\bar{C}_{mm\prime }$ obeys
the equation
\begin{eqnarray}
&\;&\lbrack (\tau ^{so}_{\varphi })^{-1}+|\Omega
_{n}|+D_{0}Q^{2}]\widetilde{C }_{00}=\tau ^{-1}\cr
&-&\frac{i}{2}v_{F}Q[\bar{C}_{-1,0}\chi _{1}^{\ast
}(\widehat{Q})+\bar{C}_{1,0}\chi _{1}(\widehat{Q})]\cr &+&O(Q^{2})
\end{eqnarray}
where  $(\tau^{so} _{\varphi })^{-1}$ is the  phase relaxation
rate contributed by spin-orbit interaction processes:
\begin{equation}
(\tau _{\varphi }^{so})^{-1}= \tau^{-1}[(\lambda^p_0)^{-1}-1].
\end{equation}
Using
\begin{equation}
\bar{C}_{\pm 1,0}=\lambda _{\pm 1}^{p}\{\bar{C}_{\pm 1,0}-\frac{%
i}{2}v_{F}Q\tau \bar{C}_{0,0}\chi _{\pm 1}(\widehat{q})\}
\end{equation}
the Cooperon is found as
\begin{eqnarray}
\bar{C}_{kk\prime }&=&\frac{1}{\tau} \frac{\gamma
_{k}^{p}\widetilde{\gamma }_{k\prime }^{p} }{|\Omega
_{m}|+D^{p}Q^{2}+\tau_{\varphi }^{-1}}\cr &\;&\cr &+& \sum_{m\neq
0}\widetilde{\lambda }_{m}^{p}\chi _{m}(\widehat{k})\chi
_{m}^{\ast }(\widehat{k}^{\prime} )\cr \widetilde{\lambda
}_{m}^{p}&=&\frac{\lambda _{m}^{p}}{1-\lambda _{m}^{p}}
\end{eqnarray}
with
\begin{eqnarray}
&\gamma _{k}^{p}&=1-\frac{i}{2}v_{F}Q\tau \sum_{m=\pm 1}
\widetilde{\lambda }_{m}^{p}\chi _{m}(\widehat{k})\chi _{m}^{\ast }(
\widehat{Q})\cr
&=&1-i\tau \sum_{m=\pm 1}\widetilde{\lambda }
_{m}^{p}\chi _{m}(\widehat{k})\langle \mathbf{Q\cdot v}_{k\prime
}\chi _{m}^{\ast }(\widehat{k}^{\prime} )\rangle
\end{eqnarray}
and
\begin{eqnarray}
\widetilde{\gamma }^p_{k}&=&1-i\tau s\sum_{m=\pm 1}
\widetilde{\lambda }^p_{m}\chi _{m}^{\ast }(\widehat{k})\langle
\mathbf{q\cdot v}_{k\prime }\chi _{m}(\widehat{k}^{\prime} )\rangle.
\end{eqnarray}
Here the diffusion coefficient $D^p$ is in general different  from
the one in the p-h channel,
\begin{equation}
D^{p}=D_{0}[1+\frac{1}{2}(\widetilde{\lambda }_{1}^{p}+\widetilde{\lambda }
_{-1}^{p})]\neq D,
\end{equation}
the difference being proportional to the spin-orbit coupling $g_{\sigma}$.

\section{Conductivity tensor in the absence of interaction}

As mentioned before, there are three mechanisms contributing to
the anomalous Hall conductivity, namely the skew scattering, the
side jump and the Berry phase mechanisms. In this section we will
write down the generic formulations for evaluating these
contributions within diagrammatic perturbation theory. The
contribution to the conductivity $\sigma _{\alpha \beta }$ will be
given in terms of a correlation function $L_{\alpha \beta }$,
defined as \cite{AGD}
\begin{equation}
\sigma _{\alpha \beta }=e^{2}\sum_{\Omega \rightarrow 0}{\lim }
\frac{1}{i\Omega _{m}}L_{\alpha \beta }
\end{equation}
where  $L_{\alpha \beta }=\sum\limits_{n}L_{\alpha \beta
}^{d_{n}}$  is a sum of the different relevant diagrams $d_n$. We
will take the current to be along $x$ direction, so the
longitudinal conductivity will correspond to $\alpha=\beta=x$
while the (anomalous) Hall conductivity will be given by the
off-diagonal part $\alpha = x, \beta=y$. Note that
$\sigma_{yx}=-\sigma_{xy}$.

\subsection{Skew scattering contribution}

The skew scattering contribution to the conductivity tensor $\sigma _{\alpha
\beta }$ \ in lowest order in $1/\varepsilon _{F}\tau $ \ is given by the
bubble diagram dressed by vertex corrections given by the correlation function
\begin{equation}
L_{\alpha \beta }=T\sum_{\epsilon _{n}}\sum_{k,\sigma } G_{k\sigma
}(i\epsilon _{n})G_{k\sigma }(i\epsilon _{n}-i\Omega
_{m})v_{k\alpha }\widetilde{j}_{k\beta }^{\sigma }
\end{equation}
The energy integration over $GG$ is nonzero only if the poles are
on opposite sides of the real axis, requiring $0\leq \epsilon
_{n}\leq \Omega _{m}$ \ (we assume $\Omega _{m}>0$ ) and yields
$2\pi N_{\sigma }\tau _{\sigma }$, and the summation on $\epsilon
_{n}$ gives  $\Omega _{m}/(2\pi T) $.   Substituting
$\widetilde{j}_{k\beta }^{\sigma}$ from Eq. (3.34) into the Kubo
formula, the conductivity tensor follows as
\begin{eqnarray}
\sigma _{\alpha \beta }^{ss}=\sum_{\mathbf{\sigma }}
\frac{1}{2}v_{F}^{2}\tau _{\sigma }N_{\sigma } \left(
\begin{array}{cc} 1+\widetilde{\lambda }_1^{\prime } &
\widetilde{\lambda }_1^{\prime\prime} \\ -\widetilde{\lambda
}_1^{\prime\prime} & 1+\widetilde{\lambda }_1^{\prime }
\end{array} \right)
\end{eqnarray}
Defining the tensor of diffusion coefficients $D_{\alpha \beta }^{\sigma }$
as
\begin{eqnarray}
D_{\alpha \alpha }^{\sigma }&=&\frac{1}{2}v_{F}^{2}\tau _{\sigma
}^{tr}\cr D_{xy}^{\sigma }&=&D_{\alpha \alpha }^{\sigma
}[\widetilde{\lambda }_1^{\prime\prime}/(1+ \widetilde{\lambda
}_1^{\prime })]=-D_{yx}^{\sigma }
\end{eqnarray}
where
\begin{equation}
\tau _{\sigma }^{tr}\equiv \tau _{\sigma }(1+\widetilde{\lambda
}_1^{\prime })
\end{equation}
is the momentum relaxation time, we may write
\begin{equation}
\sigma _{\alpha \beta }^{ss}=\sum_{\mathbf{\sigma } }N_{\sigma
}D_{\alpha \beta }^{\sigma } .
\end{equation}
From the definition $\widetilde{\lambda }_{m}=\lambda _{m}/(1-\lambda _{m})$
we obtain the following identities:
\begin{eqnarray}
1+\widetilde{\lambda }_{1}&=&\frac{1}{1-\lambda _{1}}; \;\;\;
1+\widetilde{\lambda }_1^{\prime }=\frac{1-\lambda _{1}^{\prime
}}{|1-\lambda _{1}|^{2}}\cr \widetilde{\lambda
}_1^{\prime\prime}&=&\frac{\lambda _{1}^{\prime\prime}}{|1-\lambda
_{1}|^{2}}
\end{eqnarray}

\subsection{Side-jump contribution}

The side-jump contribution has been first calculated by Berger
\cite{berger}. It arises because the trajectory of a wave packet
scattered by an impurity is shifted sidewise due to the spin-orbit
interaction ("side-jump"). This effect may be calculated in a
straightforward way \cite{crepieux} by observing that the
side-jump leads to an additional term in the particle velocity due
to the spin-orbit interaction. Indeed, the quantum mechanical
velocity obtained from the Heisenberg equation of motion for the
position operator has two terms,
\begin{equation}
\mathbf{v=}\frac{d}{dt}\mathbf{r=-}i\mathbf{[r,}H_{1}]=
\frac{\mathbf{p}}{m}+\frac{1}{4m^{2}c^{2}}(\mathbf{\tau \times
\nabla }V_{dis})..
\end{equation}
The Bloch states matrix elements of $\mathbf{v}$ are given by
\begin{eqnarray}
&\;&\langle \mathbf{k\prime \sigma \prime |v|}\mathbf{k\sigma
\rangle}= \frac{\mathbf{k}}{m}\mathbf{\delta }_{\mathbf{kk}\prime
}\mathbf{\delta } _{\sigma \sigma \prime }  -i\frac{g
_{\sigma}}{2m\varepsilon _{F}} \sum_{j}V(\mathbf{k-k\prime )}\cr
&\times& e^ {i(\mathbf{k-k\prime )\cdot }
\mathbf{R}_{j}}\{\mathbf{\tau }_{\sigma \sigma \prime }\times
(\mathbf{k-k\prime )\}}.
\end{eqnarray}

For strong impurity scattering, there are six diagrams that
contribute to the current correlation function, four of type (a)
and two of type (b), shown in Figure 1.
\begin{figure}
\begin{center}
\includegraphics[angle=0,width=0.30\textheight]{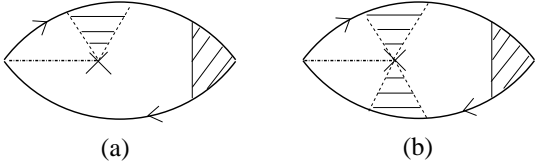}
\caption{Diagrams for side jump contributions. Solid lines are
impurity  averaged Green's functions. Shaded triangles with dashed
lines represent impurity scattering amplitudes while the dotted
line from a vertex denotes spin-orbit term in the velocity
operator. The shaded vertex represents vertex corrections to the
current density operator.}
\end{center}
\end{figure}
For example, contributions from diagrams of Figure 1 (a) and (b)
give \begin{eqnarray} L^{1a}_{xy}&=&-in_{imp}\frac{g}{\epsilon_F}T
\sum_{\mathbf{k}\mathbf{k^{\prime}}}V^2 G^+_{k}
G^+_{k^{\prime}}G^-_{k}  [\tau\mathbf{\times}
\frac{\mathbf{k}-\mathbf{k^{\prime}}}{2m}]_x\cr &\times&
f^+_{k^{\prime}k}\tilde{j}_{k_y}\cr
L^{1b}_{xy}&=&-in_{imp}\frac{g}{\epsilon_F}T
\sum_{\mathbf{k}\mathbf{k^{\prime}}}V^2  G^+_{k^{\prime}}
G^+_{k_1}G^-_{k_1} G^-_{k}  [\tau\mathbf{\times}
\frac{\mathbf{k}-\mathbf{k^{\prime}}}{2m}]_x\cr &\times&
f^+_{k^{\prime}k_1}f^-_{k_1k}\tilde{j}_{k_{1y}}
\end{eqnarray}
These were evaluated within the short range strong impurity
scattering model in Ref. [\onlinecite{WM}]. We will later use the
results reported there.

\subsection{Berry phase contribution}

In general, Berry phase contributions can arise when there is an
anomalous velocity term, as in the case of the side jump
contribution given by Eq (4.8). In principle, such terms can also
arise in the presence of a periodic potential and spin-orbit
interaction leading to finite Berry curvatures \cite{niu}. It has
been found that the intrinsic Berry curvature contributions to the
AH conductivity for bulk ferromagnetic metals can be large in
magnitude \cite{yao}. Analogous contributions for thin film
ferromagnets have not been obtained yet. Such contributions depend
on the details of the band structure and is beyond the scope of
the present work. On the other hand, the focus of the current work
is on the disorder and temperature dependence of the AH
conductivity in which the Berry contributions are qualitatively
similar to the side jump contributions (both arise from an
additional velocity term due to spin-orbit interactions).
Therefore, the effects of Berry contributions can be included in a
phenomenological way, while comparing with experiments, by
considering a larger side jump contribution to the total AH
conductivity.

\section{Interaction corrections to the conductivity}

The e-e interaction corrections to conductivity will be calculated
in first order in the screened Coulomb interaction. It may
therefore be represented as an integral over a kernel $K(q,i\omega
_{l})$ multiplied by the screened Coulomb interaction
$V_{c}(q,i\omega _{l}),$
\begin{equation}
\delta \sigma^I =T\sum_{\omega _{l}}\int
dq^{2}K(q,i\omega _{l})V_{c}(q,i\omega _{l}).
\end{equation}
Gauge invariance requires that $\delta \sigma $ should be
invariant  against an energy shift of the interaction potential,
$V(\mathbf{r)\rightarrow }V(\mathbf{r)+}C$, which only leads to a
constant term in the total Hamiltonian. In Fourier space, the
transformation is  $V(\mathbf{q)\rightarrow }V(\mathbf{q)+C\delta
(q)}$, which requires the kernel to vanish in the limit
$\mathbf{q\rightarrow }0$ \cite{kamenev}. (Even more general,
since $V(\mathbf{q)}$ is an electric potential, a gauge
transformation of the above form, but with arbitrary time
dependence $C=C(t)$ does not change the physical fields.) We will
see below that this gauge invariance, together with an additional
mirror symmetry, will impose a strong constraint on the
interaction corrections to the Hall conductivity.

\subsection{Coulomb interaction renormalized by diffusion}

The Coulomb interaction $V_{c}(q,\omega _{l})$ \ is renormalized by
diffusion processes. The bare screened interaction is given by
\begin{equation}
V_{c}(q,i\omega _{l})=V_{B}(q)/[1+V_{B}(q)\Pi (q,i\omega _{l})],
\end{equation}
where $V_{B}(q)=4\pi e^{2}/q^{2}$ \ in 3d and $V_{B}(q)=2\pi
e^{2}/q$ \ in 2d, and the polarization function is given by
\cite{lee}
\begin{equation}
\Pi (q,i\omega _{l})=\frac{dn}{d\mu }\frac{Dq^{2}}{|\omega _{l}|+Dq^{2}} .
\end{equation}
In 2d one therefore finds
\begin{equation}
V_{c}(q,i\omega _{l})=\frac{2\pi e^{2}}{q} \frac{|\omega
_{l}|+Dq^{2}}{|\omega _{l}|+Dq^{2}+DqK_{2}}\rightarrow
(\frac{dn}{d\mu })^{-1} ,
\end{equation}
in the limit $\omega _{l}=0,q\rightarrow 0$. Note that  in a
ferromagnet, an additional effective electron-electron interaction
arises by exchange of spin-wave excitations. We do not consider
this interaction here because it is small, of order
$(J/\epsilon_F)^2$ where $J$ is the exchange energy (see section
VI-B).

\subsection{Singular contributions for skew scattering}

The diagrams for the correlation functions  $L_{\alpha\beta}$
defined in (4.1) can have up to three diffusion poles
\cite{bhatt}. The gauge invariance argument presented above
suggests that the relevant contributions to $K(q,i\omega _{l})$
should have a factor of  $q^{2}$, which cancels one of the
diffusion poles. Therefore only diagrams with three diffusion
poles shown in Figure 2 contribute.
\begin{figure}
\begin{center}
\includegraphics[angle=0,width=0.28\textheight]{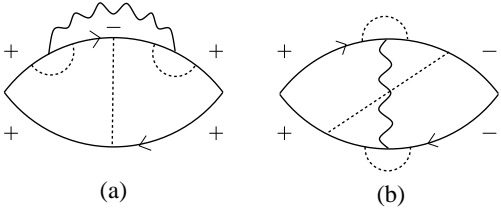}
\caption{Diagrams for interaction corrections. Solid lines are
impurity  averaged green's functions, wavy lines denote screened
coulomb interactions and dashed lines denote diffusion poles.
There are two diagrams of type (a) and two of type (b). }
\end{center}
\end{figure}
For example, contribution from diagram (a) of Figure 2 is given by
\begin{eqnarray}
L^{2a}_{\alpha\beta}&=& -T\sum_{\epsilon_n}T   \sum_{\omega_l}
\sum_{\mathbf{k,k^{\prime},
q}}G^2_{k}(\epsilon_n)G_{k-q}(\epsilon-\omega)\cr &\times&
G_{k^{\prime}-q}(\epsilon-\omega)
G_{k^{\prime}}(\epsilon_n)G_{k^{\prime}}(\epsilon_n-\Omega)
V(\mathbf{q},\omega_l)\cr &\times&[ \Theta(\epsilon)
\Theta(\epsilon-\Omega) \Theta(\omega- \epsilon)
T^{+-}_{k}(\mathbf{q},\omega)\cr &\times&
T^{-+}_{k^{\prime}}(\mathbf{-q},-\omega)
\Gamma^{+-}_{k^{\prime}k}(\mathbf{q},\omega-\Omega)\cr &+&
\Theta(-\epsilon)  \Theta(\Omega-\epsilon) \Theta(
\epsilon-\omega) T^{-+}_{k}(\mathbf{q},\omega)\cr &\times&
T^{+-}_{k^{\prime}}(\mathbf{-q},-\omega) \Gamma^{+-}_{kk^{\prime}}
(\mathbf{-q},\omega+\Omega)] \cr
&\times&v_{k\alpha}v_{k^{\prime}\beta}.
\end{eqnarray}
Using only the singular parts
\begin{eqnarray}
\Gamma^{+-}_{kk^{\prime}}(\mathbf{q},\Omega)&=&
\frac{\gamma_{k}(\mathbf{q})
\widetilde{\gamma}_{k^{\prime}}(\mathbf{q})}{|\Omega|+Dq^2}\cr
\Gamma^{-+}_{kk^{\prime}}(\mathbf{q},\omega)&=&
\Gamma^{+-}_{k^{\prime}k}(\mathbf{-q},-\omega);
\end{eqnarray}
and
\begin{eqnarray}
T^{+-}_{k}(\mathbf{q},\omega)&=& \frac{\gamma_{k}(\mathbf{q})}
{|\omega|+Dq^2}\cr  T^{-+}_{k}(\mathbf{q},\omega)&=&
\frac{\widetilde{\gamma}_{k}(\mathbf{-q},)} {|\omega|+Dq^2}
\end{eqnarray}
and defining
\begin{equation}
\mathcal{D}_q(\omega_l,\Omega_m)=\frac{V(q,\omega_l)}{(|\omega_l|
+Dq^2)^2(|\omega_l-\Omega_m|+Dq^2)}
\end{equation}
one gets
\begin{eqnarray}
L^{2a}_{\alpha\beta}&=& \sum_{\sigma}(-2\pi i N_0\tau^2)^2
\sum_{\mathbf{q}}[T\sum_{\omega_l > \Omega_m}(\omega_l -
\Omega_m)\cr &\times& \langle v_{k\alpha}\gamma_k(\mathbf{q})
\widetilde{\gamma}_k(\mathbf{q}) \xi_k(\mathbf{q})\rangle_{k}
\langle v_{k^{\prime}\beta}
\widetilde{\gamma}_{k^{\prime}}(\mathbf{q})
\gamma_{k^{\prime}}(\mathbf{q}) \xi_{k^\prime}(\mathbf{q})
\rangle_{k^{\prime}}\cr &+& T\sum_{\omega_l<0}|\omega_l| \langle
v_{k\alpha} \widetilde{\gamma}_{k}(-\mathbf{q})\gamma_{k}
(-\mathbf{q}) \xi^*_{k}(\mathbf{q})\rangle_{k} \cr &\times&
\langle v_{k^{\prime}\beta}\gamma_{k^{\prime}}
(-\mathbf{q})\widetilde{\gamma}_{k^{\prime}}
(-\mathbf{q})\xi^*_{k^\prime}(-\mathbf{q})
\rangle_{k^{\prime}}]\cr
&\times&\frac{1}{2\pi}\mathcal{D}_{\mathbf{q}}(\omega_l,\Omega_m)
\end{eqnarray}
where we have expanded the Green's functions for small $q$ and defined the factor
\begin{equation}
\xi_{k}\equiv 1-2i\tau(\mathbf{q}\cdot \mathbf{v_k}).
\end{equation}
Note that $\widetilde{\gamma}_k(-\mathbf{q},-\Omega)  =
\widetilde{\gamma}_k(\mathbf{q},\Omega)$ The leading terms in $q$
are the linear in $q$ terms in the products
$\gamma\widetilde{\gamma}\xi$:
\begin{eqnarray}
&\;&\gamma_k(\mathbf{\pm q}) \widetilde{\gamma}_k(\mathbf{\pm q})
\xi_k(\mathbf{q})  = 1\mp 2i\tau(\mathbf{q}\cdot\mathbf{v_k})\cr
&\mp& \frac{i}{2}v_F\tau \sum_{m=\pm 1}[\widetilde{\lambda}_m
+\widetilde{\lambda}^*_m] \chi_m(\widehat{k})q_{-m}
\end{eqnarray}
The $\widetilde{\lambda}$'s combine to
$\widetilde{\lambda}^{\prime}_m  =
\widetilde{\lambda}^{\prime}_{-m}$, which may be pulled in front
of the $m$-summation. Observe that
\begin{equation}
v_F\sum_{m=\pm 1}\chi_m(\widehat{k})q_{-m}=2(\mathbf{q}\cdot\mathbf{v_k}).
\end{equation}
Therefore quite generally,
\begin{equation}
\langle v_{kx}\gamma_k(\mathbf{q})
\widetilde{\gamma}_k(\mathbf{q})  \xi_k(\mathbf{q})\rangle_{k} =
-iv^2_F\tau q_x(1+\widetilde{\lambda}_1^{\prime}).
\end{equation}

\subsection{Corrections to longitudinal conductivity within skew scattering model}

For contributions from diagram (a) of Figure 2 to the longitudinal
conductivity, each of the two angular averages (in each term) in
Eq. (5.9) with $\alpha=\beta=x$ gives a factor proportional to
$q_x$ (see Eq. (5.13)), the product yielding $q^2_x$. Diagram (b)
also has the same combination. This yields, for the four diagrams
(a), (a$^{\prime}$), (b) and (b$^{\prime}$) the total contribution
($L^{2a}_{xx} =L^{2a^{\prime}}_{xx}$; $L^{2b}_{xx}
=L^{2b^{\prime}}_{xx}$):
\begin{eqnarray}
L^{2a+2a^{\prime}+2b+2b^{\prime}}_{xx} &=& \frac{1}{2\pi}
\sum_{\sigma}(2\pi N_{\sigma}\tau^2)^2(v_F^2\tau)^2
(1+\widetilde{\lambda}_1^{\prime})^2\cr &\times&
\sum_{\mathbf{q}}q^2  \Psi(q,\Omega_m),
\end{eqnarray}
where we have defined
\begin{eqnarray}
\Psi(q,\Omega_m) &=& T \sum_{\omega_l>0} \omega_l
\left[\mathcal{D}(-\omega_l,\Omega_m)-
\mathcal{D}(-\omega_l-\Omega_m,\Omega_m)\right]\cr &=& T
\left[\sum_{0<\omega_l<\Omega_m}\omega_l +
\sum_{\omega_l>\Omega_m} \Omega_m\right]\cr &\;&\cr &\times&
\mathcal{D}(-\omega_l,\Omega_m).
\end{eqnarray}
The sum over $q$ converted to an integral yields
\begin{equation}
\sum_{\mathbf{q}}q^2 \Psi(q,\Omega_m)= \frac{1}{4\pi}
\frac{e^2}{D^2\kappa}\Omega (1+\ln\frac{\omega_c}{2\pi T}),
\end{equation}
where $\kappa \equiv 2\pi e^2\sum_{\sigma}N_{\sigma}$  is the
screening length.

The exchange interaction correction to the longitudinal
conductivity is then given by
\begin{equation}
\delta\sigma^{ex}_{xx}=\frac{e^2}{\Omega_m}L_{xx} =
-\frac{e^2}{2\pi^2}\ln\frac{\omega_c}{T},
\end{equation}
where we used $D_{\sigma}=D_{0\sigma}
(1+\widetilde{\lambda}^{\prime}_{1\sigma})$. Note that the
correction $\delta\sigma_{xx}$ is independent of scattering
strength.

\subsection{Corrections to Hall conductivity within skew scattering model}

For $\alpha=x$, $\beta=y$, the two angular averages in (5.9) are
proportional to $q_x$ and $q_y$, respectively, so that the angular
$q$-integral yields zero. This is true for all four diagrams (a),
(a$^{\prime}$), (b) and  b$^{\prime}$. Thus the total correction
to the Hall conductivity $L_{xy}$ within the skew scattering model
is zero. Note that the results are true for arbitrary strength as
well as finite range and anisotropy of the impurity scattering.

Note that the result that the angular average (5.13) is
proportional  to $q_x$ is a special consequence of the fact that
(5.9) contains the combination $\gamma_k\widetilde{\gamma}_k $.
This particular combination is proportional to $\mathbf{q}\cdot
\mathbf{v_k}$ as shown in (5.12), which results in (5.13). This is
true for the class of diagrams considered here.  This leads to the
obvious question if there are other diagrams where the angular
average is over a different combination of $\gamma_k $'s leading
to a non-zero contribution to $L_{xy}$. It turns out that indeed
there are such terms with less than three diffusion poles, but
that there is a deeper reason why the {\it total} interaction
correction to the Hall conductivity must always vanish in the
first order in Coulomb interaction. In this case, the interaction
correction has the form (5.1) and the kernel must be proportional
to $q^2$ as mentioned before. In addition, we have the following
symmetry properties for the Hall conductivity with respect to a
sign change of the magnetization (magnetic field) and a mirror
reflection from the $yz$-plane $x\rightarrow -x$ (or from the
$xz$-plane $y\rightarrow -y$) which follow from the invariance of
the Hamiltonian under a simultaneous transformation $B \rightarrow
-B$ and $x \rightarrow -x$ (or $y \rightarrow -y$):
\begin{eqnarray}
\sigma_{xy}(B) &=& -\sigma_{xy}(-B) \cr
\sigma_{xy}(B;x) &=& \sigma_{xy}(-B;-x) = -\sigma_{xy}(B;-x)
\end{eqnarray}
which means that the Kernel must be proportional to $q_xq_y$ to
preserve the mirror symmetry. Thus, even though individual
diagrams do contribute, the total sum of all diagrams of a given
class must cancel to yield vanishing contribution to the Hall
conductivity. Note that the above argument remains valid for the
side jump contributions as well. Therefore we have, quite
generally,
\begin{equation}
\delta\sigma^{I}_{xy} =0.
\end{equation}
This generalizes the results of LW where this result was first
obtained within a skew scattering model with short range and weak
impurity scattering.

Note that the above arguments do not imply that the weak
localization correction to the Hall conductivity must also vanish,
because the WL contributions do not have the form Eq, (5.1) and
the gauge invariance arguments do not apply.

\subsection{Corrections to conductivity within side jump model}

We have already argued that the e-e interaction corrections to the
Hall conductivity due to side jump scattering must vanish on very
general symmetry grounds. The corresponding corrections to the
longitudinal conductivity are of course finite. However, these
contributions are proportional to the spin-orbit coupling, and
therefore are much smaller than the corrections due to normal
scattering obtained above. We will therefore neglect such
contributions.

\subsection{Hartree terms}

Eq. (5.17) should be corrected by including diagrams of the
Hartree  type. This leads to the total interaction correction in
2d \cite{altshuler}
\begin{equation}
\delta\sigma^{I}_{xx}=
-\frac{e^2}{2\pi^2}(1-\frac{3}{4}\widetilde{F}_{\sigma})
\ln\frac{\omega_c}{T},
\end{equation}
where
\begin{equation}
\widetilde{F}_{\sigma} = 8(1+F/2)\ln(1+F/2)/F -4
\end{equation}
and
\begin{equation}
F= \frac{1}{v(q=0)}\int\frac{d\theta}{2\pi} v(q=2k_F\sin\theta/2).
\end{equation}
As we will discuss later, experiments suggest an approximate
cancellation between the exchange and Hartree terms, which will
imply that the quantity
\begin{equation}
h_{xx} \equiv (1-\frac{3}{4}\widetilde{F})
\end{equation}
can be very small.

\section{Weak localization correction to conductivity}

As pointed out before, the weak localization contributions can
not be written as an integral over a kernel, as in (5.1) for the
Coulomb interaction. Therefore, although the mirror symmetry is
still preserved, the total contribution to the Hall conductivity
need not be zero.

\subsection{Cooperon contributions}
The weak localization correction to the current-current correlator
is obtained from diagrams shown in Figure 3, with one Cooperon
propagator  connecting the upper and lower line of the
conductivity bubble.
\begin{figure}
\begin{center}
\includegraphics[angle=0,width=0.25\textheight]{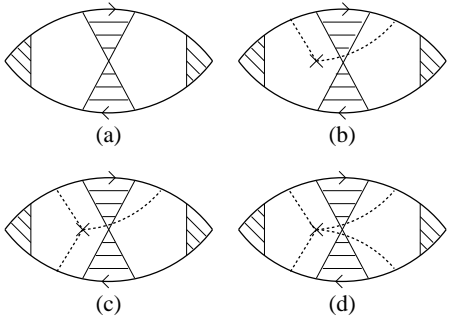}
\caption{Diagrams for weak localization corrections. Solid lines
are  impurity averaged green's functions, broken lines are
impurity scattering amplitudes. Shaded cross is the cooperon and
shaded vertices are vertex corrections to the current density
operator. There are two diagrams of type (b) and four diagrams of
type (c)}
\end{center}
\end{figure}
The frequency arguments of the upper (particle) line and the lower
(hole) line have opposite signs. The current vertices are dressed.
For example, the contribution of diagram  (a) of Figure 3 to the
current correlation function is
\begin{eqnarray}
L_{\alpha \beta }^{3a}&=&\sum_{\mathbf{\sigma }}T\sum_{ \epsilon
_{n}}\sum_{\mathbf{k,k\prime ,Q}}G_{k \sigma }(i\epsilon
_{n})G_{k\sigma }(i\epsilon _{n}-i\Omega _{m})\cr &\times&
G_{k\prime \sigma }(i\epsilon _{n})G_{k\prime \sigma }(i\epsilon
_{n}-i\Omega _{m})j_{k\alpha }^{\sigma }\widetilde{j}_{k\prime
\beta }^{\sigma } \cr &\times& (2\pi N_{\sigma }\tau _{\sigma
})^{-1}\bar{C}_{kk\prime }(\mathbf{Q};i\epsilon _{n},i\Omega _{m})
\end{eqnarray}
Here the momentum $\mathbf{Q=k+k^{\prime} }$ can be taken  to be
small, as for $Q\rightarrow 0$ \ the Cooperon is strongly peaked.
Consequently one may take $\mathbf{k^{\prime} \approx -k}$ in the
arguments of the Green's functions and of the current vertex,
i.e. $\ \widetilde{j} _{k\prime \beta }^{\sigma }\approx
-\widetilde{j}_{k\beta }^{\sigma }.$ Then
\begin{eqnarray}
L_{\alpha \beta }^{3a}&=&-(\Omega _{m}/2\pi )\sum_{\mathbf{\sigma
}}(4\pi N_{\sigma }\tau _{\sigma }^{3})(2\pi N_{\sigma }\tau
_{\sigma })^{-1} \cr &\times& \langle j_{k\alpha }^{\sigma }
\widetilde{j}_{k\beta }^{\sigma
}\rangle_{k}\sum_{\mathbf{Q}}\bar{C}_{k,-k}(\mathbf{Q})
\end{eqnarray}
The Cooperon contribution is given by
\begin{eqnarray}
\Phi &\equiv& \sum_{\mathbf{Q}}\bar{C}_{k,-k}(\mathbf{Q})\cr &=&
\int_{0}^{Q_{c}} \frac{Q dQ}{2\pi }\frac{1/\tau }{|\Omega
_{m}|+D^{p}Q^{2}+\tau _{\varphi }^{-1}}\cr &=&(4\pi \tau _{\sigma
}D^{p})^{-1}\ln (\tau _{\varphi }/\tau _{\sigma})
\end{eqnarray}
leading to a logarithmic temperature dependence through  $\tau
_{\varphi }(T)$. Similarly, contributions from the two diagrams of
type (b) can be evaluated to give
\begin{eqnarray}
L_{\alpha \beta }^{3b}&=&n_{imp}\sum_{\mathbf{\sigma }}T
\sum_{\epsilon _{n}}\{\sum_{\mathbf{k}}[G_{k \sigma }(i\epsilon
_{n})]^{2}G_{k\sigma}(i\epsilon _{n}-i\Omega _{m})\}^{2}\cr
&\times&j_{k\alpha }^{\sigma }\widetilde{j}_{k\beta }^{\sigma
}f_{k,-k\prime \sigma }^{+}f_{-k,k\prime \sigma }^{+}\Phi\cr
&=&n_{imp}\frac{\Omega _{m}}{2\pi }\sum_{\mathbf{\sigma }}(-2\pi
iN_{\sigma }\tau _{\sigma }^{2})^{2}(2\pi N_{\sigma }\tau _{\sigma
})^{-1}(\pi N_{\sigma })^{-2}\cr &\times&\langle j_{k\alpha
}^{\sigma }\widetilde{j} _{k\beta }^{\sigma }\bar{f}_{k,-k\prime
\sigma }^{+}\bar{f} _{-k,k\prime \sigma }^{+}\rangle _{k}\Phi\cr
L_{\alpha \beta }^{3b^{\prime }}&=&n_{imp}\frac{\Omega _{m}}{2\pi
}\sum_{ \mathbf{\sigma }}(2\pi iN_{\sigma }\tau _{\sigma
}^{2})^{2}(2\pi N_{\sigma }\tau _{\sigma })^{-1}(\pi N_{\sigma
})^{-2}\cr &\times&\langle j_{k\alpha }^{\sigma
}\widetilde{j}_{k\beta }^{\sigma }\bar{f}_{k\prime ,-k\sigma
}^{-}\bar{f}_{-k\prime ,k\sigma }^{-}\rangle _{k}\Phi
\end{eqnarray}
so that
\begin{eqnarray}
&\;&L_{\alpha \beta }^{3b+3b^{\prime }}=n_{imp}\frac{\Omega
_{m}}{2\pi } \sum_{\mathbf{\sigma }}(-2\pi iN_{\sigma }\tau
_{\sigma }^{2})^{2}(2\pi N_{\sigma }\tau _{\sigma })^{-1}\cr
&\times&(\pi N_{\sigma})^{-2} (v_{F}^{2}\gamma _{\sigma
})^{-1}\Phi\cr &\times&\langle j_{k\alpha }^{\sigma
}\widetilde{j}_{k^{\prime}\beta }^{\sigma }[\bar{f}_{k,-k\prime
\sigma }^{+}\bar{f}_{-k,k\prime \sigma }^{+}+\bar{f}_{k\prime
,-k\sigma }^{-}\bar{f}_{-k\prime ,k\sigma }^{-}]\rangle _{k}.
\end{eqnarray}
In a similar fashion, the total contributions from all diagrams can then be written as
\begin{eqnarray}
L^{WL}_{\alpha\beta}&=&-\frac{\Omega _{m}}{(4\pi ^2}\sum_{\mathbf{
\sigma }}(D_{\sigma }/D^p_{\sigma})J^{\alpha \beta }\ln (\tau _{\varphi
}/\tau _{\sigma })\cr
J^{\alpha \beta }&=&J_{1}^{\alpha \beta }+J_{2}^{\alpha \beta
}+4iJ_{3}^{\alpha \beta }-4J_{5}^{\alpha \beta }
\end{eqnarray}
where
\begin{eqnarray}
J^{\alpha\beta}_1&=&\frac{2}{v^2_{F\sigma}}\langle j_{k\alpha
}^{\sigma }\widetilde{j}_{k\beta }^{\sigma }\rangle \cr
J_{2}^{\alpha \beta }&=&(v_{F}^{2}\gamma _{\sigma })^{-1}\langle
j_{k\alpha }^{\sigma }\widetilde{j}_{k^{\prime}\beta }^{\sigma
}[\bar{f}_{k,-k\prime \sigma }^{+}\bar{f}_{-k,k\prime \sigma
}^{+}\cr &+&\bar{f}_{k\prime ,-k\sigma }^{-}\bar{f}_{-k\prime
,k\sigma }^{-}\rangle _{k}\cr J_{3}^{\alpha \beta
}&=&(v_{F}^{2}\gamma _{\sigma })^{-1}\langle j_{k\alpha }^{\sigma
}\widetilde{j}_{k^{\prime}\beta }^{\sigma }[\bar{f}_{k,-k\prime
\sigma }^{+}\bar{f}_{-k_{1},k\prime \sigma }^{+}\bar{f}
_{k_{1},k\sigma }^{-}\cr &-&\bar{f}_{-k\prime ,k\sigma
}^{-}\bar{f} _{k\prime ,-k_{1}\sigma }^{-}\bar{f}_{k,k_{1}\sigma
}^{+}]\rangle _{k,k\prime ,k_{1}}\cr J_{5}^{\alpha \beta
}&=&(v_{F}^{2}\gamma _{\sigma })^{-1}\langle j_{k\alpha }^{\sigma
}\widetilde{j}_{k^{\prime}\beta }^{\sigma }\bar{f} _{k,k_{2}\sigma
}^{+}\bar{f}_{-k_{1},k\prime \sigma }^{+}\cr &\times&\bar{f}
_{k\prime ,-k_{2}\sigma }^{-}\bar{f}_{k_{1},k\sigma }^{-}\rangle
_{k,k\prime ,k_{1},k_{2}}.
\end{eqnarray}
Here $J^{\alpha\beta}_1$ corresponds to contribution  from diagram
(a) of Figure 3, $J^{\alpha\beta}_2$ is a sum of contributions
from the two diagrams of type (b), $J^{\alpha\beta}_3$ is a sum of
contributions from two diagrams of type (c)  (the other two of
type (c) gives $J^{\alpha\beta}_4 = J^{\alpha\beta}_3$) and
$J^{\alpha\beta}_5$ is a contribution from diagram (d). In the
above, we have used the relation  $(n_{imp}/\pi N_{\sigma
})=1/(2\gamma _{\sigma }\tau _{\sigma })$.

\subsection{Phase relaxation rate}

The Cooperon contribution depends on the phase  relaxation rate
$\tau^{-1}_{\varphi}$, which grows linearly with temperature $T$.
In general, this may be cut off by spin-flip scattering $\tau_s$,
by spin-orbit scattering $\tau_{so}$, or by a magnetic field
characterized by $\omega_H$, all of which are independent of
temperature. Therefore, a logarithmic temperature dependence in
the conductivity requires that the phase relaxation rate satisfies
the inequality
\begin{equation}
\rm{max} (1/\tau_s, 1/\tau_{so}, \omega_H) \ll 1/\tau_{\varphi} \ll 1/\tau_{tr}.
\end{equation}
The contribution to $\tau_{\phi}$ from e-e interaction is given by
\begin{equation}
1/\tau_{\varphi}= \frac{T}{\epsilon_F\tau_{tr}} \ln\frac{\epsilon_F\tau_{tr}}{2}.
\end{equation}
This is typically too small to satisfy the above inequality in
thin ferromagnetic films where in particular the internal magnetic
field $B_{in}$ can be estimated to give rise to
$\omega_H=4(\epsilon_F\tau_{tr})(eB_{in}/m^*c)$ which can be
large. A much larger contribution is obtained from scattering off
spin-waves in such systems \cite{tatara}, which is given by
\begin{equation}
1/\tau_{\varphi}=4\pi T \frac{J^2}{\epsilon_F\Delta_g},
\end{equation}
where $J$ is the exchange energy of the $s$-electrons and
$\Delta_g$ is the spin-wave gap. As estimated in  Ref.
[\onlinecite{mitra}], with this contribution to the phase
relaxation rate, the inequality (6.8) can be satisfied within
experimentally accessible disorder and temperature ranges where
the WL effects can be observed.

\section{Strong short range impurity scattering}

The results of the previous section can in  principle be used to
obtain the weak localization corrections to both longitudinal and
Hall conductivities. However, The algebra gets fairly involved
without contributing extra insight into the problem. Since higher
angular momentum components are expected to be smaller, we will
consider the dominant contribution that arises from a short range
impurity model and show in the Appendix how effects of finite
range anisotropic scattering can be included within model
calculations. On the other hand, we will keep the calculations
valid for arbitrary strength of the impurity scattering.

\subsection{Scattering amplitude, relaxation rate and particle-hole and particle-particle propagators}

These were already obtained for short range strong impurity
scatterings in Ref. [\onlinecite{WM}] and we will simply quote the
results. The scattering amplitude is given by
\begin{equation}
\bar{f}_{k\sigma ,k\prime \sigma \prime
}=\frac{\widetilde{w}}{\sqrt{w }}-i\tau _{\sigma \sigma
}^{z}(\widehat{k}\times \widehat{k}\prime )\frac{2
\widetilde{u}}{\sqrt{u}}-is_{\omega
_{n}}[\widetilde{w}+2\widetilde{u}( \widehat{k}\cdot
\widehat{k}\prime )].
\end{equation}
Here we defined  $\widetilde{w}=w/(1+w)$,  and
$\widetilde{u}=u/(1+u)$, where  $w=(\pi N_{\sigma }V)^{2}$  and
$u=(g _{\sigma}/2)^{2}w$ , and all quantities depend on the spin
orientation $\sigma $ (suppressed here and in the following,
except in the final expressions involving spin summation). In
terms of the angular momentum components  of $\bar{f}$ defined in
(3.9), $\bar{f}^s_m$, we have from (7.1):
\begin{eqnarray}
\bar{f}_{kk\prime }^{s}&=&\bar{f}_{0}^{s}+\bar{f}_{1}^{s}
\widehat{k}_{+}\widehat{k}_{-}^{\prime
}+\bar{f}_{-1}^{s}\widehat{k} _{-}\widehat{k}_{+}^{\prime };
\;\;\; \bar{f}_{0}^{s}=\frac{
\widetilde{w}}{\sqrt{w}}-is\widetilde{w}\cr \bar{f}_{\pm
1}^{s}&=&-is\widetilde{u}\pm \tau _{\sigma \sigma }^{z}
\frac{\widetilde{u}}{\sqrt{u}}; \;\;\; \bar{f}_{m}^{s}= 0, \;\;\;
|m| > 1.
\end{eqnarray}

Using Eq.(7.2), the single particle relaxation rate given by Eq. (3.10) becomes
\begin{equation}
\frac{1}{2\tau_{\sigma}} =\frac{n_{imp}}{\pi N_{\sigma
}}(\widetilde{w}+2\widetilde{u}).
\end{equation}
One observes that $\frac{1}{2\tau _{\sigma }}$ \ is proportional
to the Fermi energy, the average number of impurities per electron
and the dimensionless factor \ $(\widetilde{w}+2\widetilde{u})$,
expressing the effective scattering strength per impurity.
Eigenvalues of the particle-hole scattering amplitude
$\bar{t}^{+-}_{kk^{\prime}}$ are obtained to be
\begin{eqnarray}
\lambda _{0} &=& 1; \;\;\; \lambda _{-m}=\lambda _{m}^{\ast },\cr
\lambda_{1} &=& 2\widetilde{w}\widetilde{u}(\widetilde{w}+2\widetilde{u}
)^{-1}(1+is\frac{1}{\sqrt{u}}\tau _{\sigma \sigma }^{z}) \cr
\lambda_{2} &=& \frac{\widetilde{u}^{2}}{u}(\widetilde{w}+2\widetilde{u}
)^{-1}(u-1+2is\sqrt{u}\tau _{\sigma \sigma }^{z})
 \end{eqnarray}
while for $\bar{t}^{++}_{kk^{\prime}}$ one obtains (with $
\bar{t}_{kk\prime }^{ss}\equiv \sum_{m}\xi _{m}\chi
_{m}(\widehat{k})\chi _{m}^{\ast }(\widehat{k} \prime )$)
\begin{eqnarray}
\xi_{0} &=& (\widetilde{w}+2\widetilde{u})^{-1}[\frac{\widetilde{w}}{1+w}
(1-w-2is\sqrt{w})+2\widetilde{u}\frac{1-u}{1+u}] \cr
\xi_{1} &=& -2\widetilde{w}\widetilde{u}(\widetilde{w}+2\widetilde{u}%
)^{-1}(1+is\frac{1}{\sqrt{w}}) \cr
\xi _{2} &=& -\widetilde{u}(\widetilde{w}+2\widetilde{u})^{-1}
\end{eqnarray}
It may be shown that \ $\Delta \ \bar{t}_{kk\prime }(q)$ defined
in (3.13) gives rise to small corrections to the diffusion
coefficient, of order $(1/\varepsilon _{F}\tau )$  and hence may
be dropped.

Eigenvalues of the particle-particle scattering amplitude
$\bar{t}^{p,+-}_{kk^{\prime}}$ are obtained to be
\begin{eqnarray}
\lambda _{0}^{p}&=&[\widetilde{w}-2\widetilde{u}(1-2\widetilde{u})]/(
\widetilde{w}+2\widetilde{u})\cr
\lambda _{\pm 1}^{p}&=&(2\widetilde{w}\widetilde{u}\pm 2\frac{\widetilde{w}}{
\sqrt{w}}\frac{\widetilde{u}}{\sqrt{u}}\tau _{\sigma \sigma }^{z})/(
\widetilde{w}+2\widetilde{u})\cr
\lambda _{\pm 2}^{p}&=&\widetilde{u}/(\widetilde{w}+2\widetilde{u})
\end{eqnarray}
We observe that $\lambda _{0}^{p}\neq 1$ \ if skew scattering is present, as
it violates time reversal symmetry.

The phase relaxation rate $(\tau^{so} _{\varphi })^{-1}$  defined
in Eq. (3.42) is given by
\begin{equation}
(\tau^{so} _{\varphi })^{-1}=\tau ^{-1}4\widetilde{u}(1-\widetilde{u})/[\widetilde{
w}-2\widetilde{u}(1-2\widetilde{u})]
\end{equation}
which is positive for not too large
spin-orbit scattering,  $u\lesssim w/2,$ or  $g_{\sigma }\lesssim 1.$

\subsection{Hall conductivity}

The conductivity tensor due to skew scattering was already
evaluated in section IV.A for general strong finite range impurity
scattering in terms of the eigenvalues of the particle-hole
propagator $\lambda$. In particular, it gives
\begin{equation}
\frac{\sigma^{ss}_{xy}}{\sigma^{ss}_{xx}} =
\frac{\lambda^{\prime\prime}_1}{1-\lambda^{\prime}_1}
\end{equation}
For short range scattering, Eq. (7.4) gives explicit  expressions
for the eigenvalues in terms of the scattering potentials. The
side jump contribution was already evaluated in Ref.
\onlinecite{WM} and we quote the result:
\begin{equation}
\sigma^{sj}_{xy}=\frac{e^2}{2\pi}\sum_{\sigma}
\tau^z_{\sigma\sigma}g_{\sigma}
\frac{\tilde{w}}{\tilde{w}+2\tilde{u}}
\frac{(1+\tilde{\lambda}^{\prime}_1)}{1+u}
\end{equation}
Using (7.4), this yields, in the small $u\ll w\ll 1$ limit,
\begin{equation}
\sigma^{sj}_{xy}=\frac{e^2}{2\pi}\sum_{\sigma}
\tau^z_{\sigma\sigma}g_{\sigma} \frac{1}{1-\lambda^{\prime}_1}
\end{equation}

\subsection{Weak localization correction}

Evaluation of  $J^{\alpha \beta }$ defined in section VI (Eqs.
(6.6), (6.7)) in the present short-range (but arbitrary scattering
strength) model gives
\begin{eqnarray}
J_{1}^{xx}&=& (1+\widetilde{\lambda}^{\prime}_1)^2
-(\widetilde{\lambda}^{\prime\prime}_1)^2; \;\;\; J_{1}^{xy}= 2
\tilde{\lambda}^{\prime\prime }_1(1+\tilde{\lambda}^{\prime}_1)\cr
J_{2}^{xx} &=&[\lambda _{1}^{\prime }J_{1}^{xx}-\lambda
_{1}^{\prime\prime }J_{1}^{xy}]; \;\;\; J_{2}^{xy}=[\lambda
_{1}^{\prime\prime }J_{1}^{xx}+\lambda _{1}^{\prime
}J_{1}^{xy}]\cr J_{3}^{xx} &=&\frac{i}{2}\{2\widetilde{u}\lambda
_{1}^{\prime }J_{1}^{xx}-(2 \widetilde{u}+1)\lambda
_{1}^{\prime\prime }J_{1}^{xy}\}\cr
J_{3}^{xy}&=&\frac{i}{2}\{(2\widetilde{u}+1)\lambda
_{1}^{\prime\prime }J_{1}^{xx}+ 2\widetilde{u}\lambda _{1}^{\prime
}J_{1}^{xy}\}\cr J_{5}^{xx}&=&
-\frac{1}{2}\{(2\widetilde{u}-1)\lambda _{1}^{\prime }J_{1}^{xx}-
2\widetilde{u}\lambda _{1}^{\prime\prime }J_{1}^{xy}\}\cr
J_{5}^{xy}&=&-\frac{1}{2}\{2\widetilde{u}\lambda
_{1}^{\prime\prime }J_{1}^{xx}+ (2\widetilde{u}-1)\lambda
_{1}^{\prime }J_{1}^{xy}\}\cr iJ_{3}^{\alpha \beta
}&-&J_{5}^{\alpha \beta }=-\frac{1}{2}J^{\alpha\beta}_2; \;\;\;
J^{\alpha\beta} = J^{\alpha\beta}_1-J^{\alpha\beta}_2.
\end{eqnarray}
We may combine this into the compact expression
\begin{equation}
J^{xx}=Re\{\Lambda \}; \;\;\; J^{xy}=Im\{\Lambda \}; \;\;\;
\Lambda =\frac{1}{1-\lambda_1}
\end{equation}

Note that the final result for $J^{\alpha\beta}$ contains
detailed effects of the potentials only through the eigenvalues
$\lambda$. This suggests that the results may be more general than
the short range potentials used in the calculations. Also, as we
will show in the Appendix, $\lambda _{1}^{\prime }$ may approach
unity in the limit of extreme forward scattering.

In any case, for the short range impurity scattering model
considered above, we then have contributions from weak
localization corrections given by
\begin{eqnarray}
\delta\sigma^{WL}_{xx} &=& -\frac{e^2}{4\pi^2}\sum_{\mathbf{
\sigma }}(D_{\sigma }/D^{p})\ln (\tau _{\varphi}/\tau _{\sigma
})\cr \frac{\delta\sigma^{WL}_{xy}}{\delta\sigma^{WL}_{xx}} &=&
\frac{Im( \Lambda)}{Re (\Lambda)}=
\frac{\lambda^{\prime\prime}_1}{1-\lambda_1^{\prime}}.
\end{eqnarray}

\section{Comparison with Experiments}

Experiments measure the longitudinal and Hall resistances
$R_{\alpha\beta}$ as functions of both sheet resistance and
temperature. In order to compare, we obtain the normalized
relative conductances defined as
\begin{equation}
\Delta^N\sigma_{\alpha\beta} \equiv \frac{1}{L_{00}R_0}
\frac{\delta\sigma_{\alpha\beta}}{\sigma_{\alpha\beta}}
\end{equation}
where $L_{00} \equiv e^2/2\pi^2$ and $R_0=1/\sigma_{xx}$ As shown
above, a logarithmic temperature dependence in  these quantities
can arise either from interaction corrections or from weak
localization corrections. However, although two separate groups
have seen such logarithmic temperature dependences
\cite{mitra,BY}, the prefactors seem to be more universal for
$\Delta^N\sigma_{xx}$, independent of sheet resistance $R_0$ or
sample preparation for a range of $ R_0 $,  but clearly disorder
and sample dependent for $\Delta^N\sigma_{xy}$ in the same range
of $R_0$.  In this section we collect all our results above to
obtain the total contribution to $\Delta^N\sigma_{\alpha\beta}$
from all possible mechanisms considered above. As used in the
text, superscripts $ss$ and $sj$ will refer to the skew scattering
and side jump mechanisms, and $I$ and $WL$ will refer to the
interaction and weak localization corrections, respectively. While
the results for $\sigma^{ss}_{\alpha\beta}$ and
$\delta\sigma^I_{xy}$ are valid for finite range strong impurity
scatterings, others are evaluated within a short range strong
impurity scattering model. We have also assumed that the
spin-orbit coupling is weak.

The conductivities due to skew and side jump scatterings are
\begin{eqnarray}
\sigma^{ss}_{xx}&=&\sum_{\sigma}
\frac{1}{2}v^2_{F\sigma}N_{\sigma} \tau_{tr}; \;\;\;
\sigma^{sj}_{xx} \ll \sigma^{ss}_{xx}\cr \sigma^{ss}_{xy}&=&
\sigma^{ss}_{xx}\frac{\lambda^{\prime\prime}_1}
{1-\lambda^{\prime}_1}\cr \sigma^{sj}_{xy}&=&
\frac{e^2}{2\pi}\sum_{\sigma} \tau^z_{\sigma\sigma}
g_{\sigma}\frac{ (1-\lambda^{\prime}_1)}{|1-\lambda_1|^2}
\end{eqnarray}
Quantum corrections to the conductivities due to Coulomb
interaction and weak localization effects leading to a logarithmic
temperature dependence are
\begin{eqnarray}
&\;&\delta\sigma^{ss,I}_{xx}= L_{00} h_{xx} \ln(T\tau); \;\;\;
\delta\sigma^{ss,WL}_{xx} = L_{00}\ln (T\tau)\cr
&\;&\delta\sigma^{ss,I}_{xy}= 0; \;\;\; \delta\sigma^{ss,WL}_{xy}
= \delta\sigma^{ss,WL}_{xx}
\frac{\lambda^{\prime\prime}_1}{1-\lambda^{\prime}_1}\cr
&\;&\delta\sigma^{sj,I}_{xy} =0; \;\;\; \delta\sigma^{sj,I}_{xx}
\ll  \delta\sigma^{ss,I}_{xx}\cr
&\;&\delta\sigma^{sj,WL}_{\alpha\beta} \ll
\delta\sigma^{ss,WL}_{xy}
\end{eqnarray}
The total conductivities and quantum corrections are simply
\begin{eqnarray}
\sigma_{xx} &=& \sigma^{ss}_{xx}; \;\;\; \sigma_{xy} =
\sigma^{ss}_{xy} + \sigma^{sj}_{xy}\cr \delta\sigma_{xx} &=&
\delta\sigma^{ss,I}_{xx} +  \delta\sigma^{ss,WL}_{xx}; \;\;\;
\delta\sigma_{xy} =  \delta\sigma^{WL}_{xy};
\end{eqnarray}
Using these results, we obtain
\begin{eqnarray}
\Delta^N\sigma_{xx} &=& \frac{\sigma^{ss}_{xx}}{L_{00}}
\frac{\delta\sigma^{ss,I}_{xx} +
\delta\sigma^{ss,WL}_{xx}}{\sigma^{ss}_{xx}}=(1+h_{xx})
\ln(T\tau)\cr \Delta^N\sigma_{xy} &=&
\frac{\sigma^{ss}_{xx}}{L_{00}}
\frac{\delta\sigma^{ss,WL}_{xy}}{\sigma^{ss}_{xy}
+\sigma^{sj}_{xy}}= \frac{1}{(1+r_{xy})}\ln(T\tau)
\end{eqnarray}
where $h_{xx}$ defined in Eq, (5.23) is the exchange plus Hartree
interaction contribution to the longitudinal conductivity and we
have defined
\begin{equation}
r_{xy}\equiv \frac{\sigma^{sj}_{xy}}{\sigma^{ss}_{xy}}
\end{equation}
as the ratio of side jump to skew scattering contributions to  the
Hall conductivity. Note that $r_{xy}$ is a non-universal quantity.
As shown in \cite{mitra}, all current experiments can be
understood if $h_{xx} \ll 1$ and $r_{xy}$ is sample dependent and
is allowed to vary with disorder. In particular, this means that
while the skew scattering and side jump mechanisms both contribute
to the AH conductivity, the side jump contributions to the
longitudinal conductivity as well as to the weak localization
corrections to the conductivity tensor are much smaller than the
corresponding skew scattering contributions when the spin-orbit
coupling is weak.

\section{Summary and conclusion}

We develop a systematic general formulation for  the AHE for
strong, finite range impurity scattering starting from a
microscopic model of electrons in a random potential of impurities
including spin-orbit coupling. In particular, we consider quantum
corrections to the AH conductivity, observed in different
experiments on disordered thin ferromagnetic films with apparently
different results. General symmetry arguments presented here show
that the e-e interaction corrections must vanish exactly, which
then implies that there must be weak localization corrections in
these ferromagnetic films despite the presence of large internal
magnetic fields.

Our evaluations of the WL effects within a short range but  strong
impurity scattering lead to the normalized relative conductances
given by Eq. (8.5), where the spin-orbit coupling has been assumed
to be weak. These results are consistent with all experimental
observations, where the difference between different experiments
arise due to different contributions from skew scattering vs side
jump mechanism.

In this paper we have only briefly mentioned the Berry phase
effects. A systematic study of the Berry phase contributions to
the AHE will be reported elsewhere.
\bigskip

\centerline{{\bf ACKNOWLEDGEMENTS:}}
\bigskip
We thank A. Hebard, R. Misra and P. Mitra for useful discussions
on the experimental data on Fe film. This work has been supported
by the DFG-Center for Functional Nanostructures at the Karlsruhe
Institute of Technology (KIT).

\bigskip

\centerline{{\bf APPENDIX:}}

\centerline{{\bf LONG RANGE CORRELATED POTENTIALS}}
\bigskip
For completeness, here we consider models to  incorporate possible
effects of small and large angle scattering.

\subsection{ Model of small angle scattering}

Long range correlated potentials will scatter  electrons
predominantly by a small angle $\theta <<\pi $ . A simple model is
provided by a gaussian dependence
\begin{equation}
V(\mathbf{k-k\prime })=V(\theta )=4\sqrt{\pi }V_{0}\theta
_{0}^{-1}e^{-(\theta /\theta _{0})^{2}}
\end{equation}
where $\theta _{0}<<\pi $.
The angular momentum components of $V(\theta )$ are given by
\begin{equation}
V_{m}^{ns}=\int_{0}^{\pi }\frac{d\theta }{2\pi }V(\theta
)=V_{0}e^{-m^{2}\theta _{0}^{2}/4}
\end{equation}
In the limit of weak scattering we have $\bar{f}_{m\sigma }
=\bar{V}_{m\sigma}$ and then
\begin{equation}
\gamma _{\sigma }=\sum_{m}|\bar{V}_{m\sigma }|^{2}=(\pi N_{\sigma
}V_{0})^{2}\sqrt{2\pi }/\theta _{0}.
\end{equation}
Neglecting skew scattering for the moment we find
\begin{eqnarray}
\bar{t}_{1\sigma }^{+,-}&=&\gamma _{\sigma }^{-1}(\pi N_{\sigma
}V_{0})^{2}\sum_{m}e^{ -\frac{\theta _{0}^{2}}{4}
(m^{2}+(m-1)^{2})}\cr &=&e^{ -\theta _{0}^{2}/8}.
\end{eqnarray}
It follows that  $1-\bar{t}_{1\sigma }^{+,-}\approx \theta
_{0}^{2}/8\ll 1$ and therefore the diffusion coefficient  is
enhanced by a factor
\begin{equation}
D/D_{0}=(\theta _{0}^{2}/8)^{-1} .
\end{equation}

\subsection{ Model of strong back-scattering}

It is well known, that the scattering of conduction electrons in amorphous
metals can be anomalous in the sense that the transport relaxation time is
smaller than the single particle relaxation time. This is due to the fact
that the atomic structure is characterized by finite range order. The pair
correlation function shows enhanced peaks corresponding to the nearest
neighbor, next nearest neighbor, etc. shell. In other words, the system
shows crystalline order over a certain usually short distance. As a
consequence electrons are suffering Bragg scattering by large angles. The
scattering cross section for large angles is larger than that for small
angles. Consequently the angular average of the cross section $\sigma
(\theta )$, weighted with the factor \ $(1-\cos \theta ),$ appearing in the
expression for the transport relaxation rate is larger than the uniform
average in the single particle transport rate. In the case of
polycrystalline material we expect a similar effect.

The scattering potential $V(r)$ of a crystallite or a small grain of
amorphous metal will show oscillating behavior in real space reflecting the
nearly regular arrangement of atoms, and its Fourier transform will show a
peak at a finite momentum $q=2\pi /a$ corresponding to the spatial period $a$
, which will be equal or close to the lattice constant of the crystalline
phase. The width of the peak will be determined by the range of the short
range order or the size of the crystallites. This is in contrast to a usual
impurity potential whose Fourier transform has a peak at $q=0$ and a width
corresponding to the range of the potential. In terms of the angular
momentum components $V_{l}$ of the scattering potential a peak in $V(q)$
implies that some of the $V_{l}$ will be negative. \ In particular,  the
component $\lambda _{1}$of the t-matrix \ $t_{kk\prime }$ determining the
transport relaxation rate  will be negative.

Let us consider a simple model of a crystallite of size $L$. Its scattering
potential seen by a conduction electron of the matrix (assumed to be
isotropic, as appropriate for an amorphous system) is something like
\begin{eqnarray}
V_{1}(x)&=&V_{0}\cos (2\pi x/a)\theta (L/2-|x|)\cr
&=&V_{0}S_{1}(x); \;\;\; \rm{1d} \cr
V_{2}(x,y)&=&V_{0}S_{1}(x)S_{1}(y); \;\;\;  \rm{2d}
\end{eqnarray}
The Fourier transform of $S_{1}(x)$ is given by
\begin{eqnarray}
S_{1}(k)=\frac{L}{2}\frac{[K\cos (K)\sin (\kappa )-\kappa \sin (K)\cos (\kappa
)]}{[K^{2}-\kappa ^{2}]},
\end{eqnarray}
where $ K=kL/2$, $\kappa =\pi L/a$.
$S_{1}(k)$ increases linearly with $k$ at small $k,$ has maximum at $
k\approx 2\pi /a$ and decreases as \ $1/k$ for large $k$.  We may model
this behavior by
\begin{equation}
V_{2}(k)=V_{0}\frac{kk_{0}}{k_{0}^{2}-k^{2}},
\end{equation}
where  $k_{0}=2\pi/a$. Using \ the relation of the  transferred
momentum $\mathbf{k=k}_{f}-\mathbf{k} _{i}$ to the scattering
angle $\phi $,  $k^{2}=2k_{F}^{2}(1-\cos \phi )$, where
$|\mathbf{k}_{f,i}|=k_{F}$, we get
\begin{equation}
V_{2}(\phi )=\overline{V}\frac{\sqrt{1-\cos \phi }}{\eta +\cos
\phi },
\end{equation}
where $\overline{V}=V_{0}/(k_{F}\sqrt{2})$,  $\eta
=k_{0}^{2}/2k_{F}^{2}-1$.

The angular momentum components $V_{l}$ may be calculated as
\begin{equation}
V_{l}=\int_{0}^{2\pi }\frac{d\phi }{2\pi }\cos (l\phi )V_{2}(\phi ).
\end{equation}
In particular we find
\begin{eqnarray}
V_{0} &=&\frac{2}{\pi }\overline{V}(\eta -1)^{-1/2}\arctan \sqrt{\frac{2}{\eta
-1}}>0 \cr
V_{1} &=& \frac{2}{\pi }\overline{V}\{-\frac{\eta }{\sqrt{\eta -1}}\arctan
\sqrt{\frac{2}{\eta -1}}+\sqrt{2}\}\cr
&\leq& 0
\end{eqnarray}
In the limit $\eta \rightarrow 1$  the ratio of  the  $l=1$ and $l=0$
components is given by \ $V_{1}/V_{0}=-\eta $.  We may estimate $\eta $
by assuming $Z$ electrons in a unit cell of area $a^{2}$ resulting in $
k_{F}^{2}=2\pi Z/a^{2}$ and therefore  $\eta =2\pi /Z-1$.  For  $Z\approx
2,5$  appropriate for a mixture of $Fe^{2+}$ and   $Fe^{3+}$  one finds
$\eta \approx 1.5$ and then $V_{1}/V_{0}\approx -0.6$. In the
following we will take the $V_{l}$ to be given parameters, which may be
negative.

In order to keep the calculation simple we will neglect all
angular momentum components with  $|l|\geq 2$.  Defining
dimensionless quantities $ \bar{V}_{l}=\pi N_{\sigma }V_{l}$ as
before, the dimensionless scattering amplitudes are given by
\begin{eqnarray}
\bar{f}_{0}^{s}&=&\bar{V_0}/(1+is \bar{V}_{0}); \;\;\;\bar{f}_{\pm
1,\sigma }^{s}=\bar{V}_{\pm 1,\sigma }/(1+is\bar{V}_{\pm 1,\sigma
})\cr \bar{V}_{\pm 1,\sigma }&=&\bar{V_1}\pm \sqrt{u}\tau _{\sigma
\sigma }^{z}.
\end{eqnarray}
Assuming weak spin-orbit scattering we may expand in $\sqrt{u}$:
\begin{equation}
\bar{f}_{\pm 1,\sigma }^{s}=\frac{\bar{V_1}}{1+i s \bar{V}_{1}}\pm
(1+i s \bar{V}_{1})^{2}\sqrt{u}\tau _{\sigma \sigma }^{z}
\end{equation}
The normalization factor  $\gamma _{0}$ entering the expression for the
relaxation rate is obtained as
\begin{equation}
\gamma _{0}=\frac{w}{1+w}+2\frac{w_{1}}{1+w_{1}}+O(\sqrt{u}),
\end{equation}
where $w=\bar{V_0}^{2}$,  $w_{1}=\bar{V_1}^{2}$.  The eigenvalue
$\lambda _{1}$ of \ $t_{kk\prime }$ is found as
\begin{eqnarray}
\lambda _{1}&=&\frac{1}{\gamma
_{0}}\{\frac{2\bar{V_0}\bar{V_1}(1+\bar{V_0}\bar{V_1})}{
(1+w)(1+w_{1})}\cr &+&2i\sqrt{u}\tau _{\sigma \sigma
}^{z}\bar{V_0}\frac{
\bar{V_0}(1-w_{1})-2\bar{V_1}}{(1+w)(1+w_{1})^{2}}\}
\end{eqnarray}
Analyzing this expression one finds that the largest negative
values of  $ \lambda _{1}$ are reached for weak scattering,
$\bar{V_0},\bar{V_1}\ll 1$, when
\begin{equation}
\lambda
_{1}=\frac{2\bar{V_0}}{w+w_{1}}[\bar{V_1}+i(\bar{V_0}-2\bar{V_1})\sqrt{u}\tau
_{\sigma \sigma }^{z}]
\end{equation}
The minimum of \ $\lambda _{1}^{\prime }$ is obtained  if
$\bar{V_1}/\bar{V_0}=-1/ \sqrt{2}$,  where  $\lambda _{1}^{\prime
}=-1/\sqrt{2}$.

Let us now consider diagram  $w_{2}$,  which is determined by  the
parameter  $J_{2}^{\alpha \beta }$,  given by
\begin{equation}
J_{2}^{xx}=-\gamma _{0}^{-1}\{(1+\widetilde{\lambda
}_{1})^{2}(\bar{f} _{0}^{+}\bar{f}_{+1,\sigma
}^{+}+\bar{f}_{0}^{-}\bar{f} _{-1,\sigma }^{-})+c.c.\}
\end{equation}
\begin{eqnarray}
b_{1} &\equiv& \bar{f}_{0}^{+}\bar{f}_{+1,\sigma }^{+}+\bar{f}
_{0}^{-}\bar{f}_{-1,\sigma }^{-} \cr &=&
\frac{2\bar{V_0}}{(1+w)(1+w_{1})}
\{\bar{V_1}(1-\bar{V_0}\bar{V_1})\cr &-&i\sqrt{u}\tau _{\sigma
\sigma }^{z}\frac{ \bar{V_0}(1-w_{1})+2\bar{V_1}}{(1+w_{1})}\}
\end{eqnarray}
In the weak scattering limit,  we have
\begin{equation}
\beta _{1} \equiv b_{1}/\gamma
_{0}=\frac{2\bar{V_0}}{w+w_{1}}[\bar{V_1}-i(2\bar{V_1}+\bar{V_0})
\sqrt{u}\tau _{\sigma \sigma }^{z}],
\end{equation}
which differs from \ $\lambda _{1}$ only by the sign of the term \
$\bar{V_0}$ in the imaginary part, i.e. $\beta _{1}^{\prime
}=\lambda _{1}^{\prime }$.

\end{document}